\documentclass[journal]{IEEEtran}
\usepackage{graphicx}
\usepackage{algorithm,algcompatible,amsmath}
\usepackage{bm}
\usepackage{comment}
\usepackage{amssymb}
\usepackage{xcolor}
\usepackage{subfigure} 
\usepackage{hyperref}
\graphicspath{{./Figures/}}
\newtheorem{theorem}{Theorem}[section]

\newtheorem{proposition}[theorem]{Proposition}
\usepackage{cite}
\ifCLASSINFOpdf
\else
\fi
\DeclareMathOperator{\EX}{\mathbb{E}}

\begin{document}
\title{Non-uniform Array and Frequency Spacing for Regularization-free Gridless DOA}
\author{Yifan Wu,
        Michael B. Wakin, \IEEEmembership{Fellow, IEEE} and
        Peter Gerstoft, \IEEEmembership{Fellow, IEEE}
\thanks{Yifan Wu and Peter Gerstoft are with NoiseLab, University of California, San Diego, La Jolla, CA, 92093, USA (e-mails:\{yiw062, pgerstoft\}@ucsd.edu).
}
\thanks{Michael. B. Wakin is with Colorado School of Mines, Golden, CO, 80401, USA (e-mail: mwakin@mines.edu)}
\thanks{This work is supported by NSF Grant CCF-1704204, NSF Grant CCF-2203060, and Office of Naval Research (ONR) Grant N00014-21-1-2267. }
\thanks{Manuscript received \today}}
\markboth{IEEE Transactions on Signal Processing}%
{Shell \MakeLowercase{\textit{et al.}}: Bare Demo of IEEEtran.cls for IEEE Journals}

\maketitle
\begin{abstract}
Gridless direction-of-arrival (DOA) estimation with multiple frequencies can be applied in acoustics source localization problems. We formulate this as an atomic norm minimization (ANM) problem and derive an equivalent \textit{regularization-free} semi-definite program (SDP) thereby avoiding regularization bias. The DOA is retrieved using a Vandermonde decomposition on the Toeplitz matrix obtained from the solution of the SDP. 
We also propose a fast SDP program to deal with non-uniform array and frequency spacing. For non-uniform spacings, the Toeplitz structure will not exist, but the DOA is retrieved via irregular Vandermonde decomposition (IVD), and we theoretically guarantee the existence of the IVD. We extend ANM to the multiple measurement vector (MMV) cases and derive its equivalent regularization-free SDP. Using multiple frequencies and the MMV model, we can resolve more sources than the number of physical sensors \textit{for a uniform linear array}. Numerical results demonstrate that the regularization-free framework is robust to noise and aliasing, and it overcomes the regularization bias. 
\end{abstract}

\IEEEpeerreviewmaketitle

\section{Introduction}
\IEEEPARstart{D}{irection}-of-arrival (DOA) estimation is an important topic in sensor array processing \cite{van2002optimum} that has a broad range of applications in wireless communication \cite{chen2021millidegree}, radar \cite{vasanelli2020calibration}, remote sensing, etc. Conventional DOA estimation methods (e.g. multiple signal classification (MUSIC) \cite{schmidt1986multiple}, and estimation of signal parameters via rotational invariant techniques (ESPRIT) \cite{roy1989esprit}) are mainly developed for narrowband signals. In the past few decades, some wideband DOA estimation methods have been proposed \cite{wax1984spatio, antonello2019joint, nannuru2019sparse, gemba2019, wu2023gridless, wang1985coherent, buckley1988broad, di2001waves, yoon2006tops, tang2011aliasing, gemba2017}. Recently proposed methods based on sparse recovery and a multi-frequency model \cite{nannuru2019sparse, wu2023gridless} have demonstrated superior performance in wideband DOA estimation problems. Before introducing the contributions of this paper, we review the relevant prior works.

\subsection{Related Work}
\subsubsection{Wideband DOA Estimation and Multiple Frequencies}
Wideband signal DOA estimation has been studied for decades \cite{wax1984spatio, wang1985coherent, buckley1988broad, di2001waves, yoon2006tops}. In \cite{wax1984spatio}, a subspace-based wideband DOA estimation approach, the incoherent signal subspace method (ISSM), was proposed. The coherent signal subspace method (CSSM) \cite{wang1985coherent} led to improved performance compared to ISSM. A broadband spatial-spectrum estimation approach \cite{buckley1988broad} overcame the peak bias and source spectral content sensitivity from CSSM. Variants of CSSM, such as the weighted average of signal subspaces method \cite{di2001waves} and the test of orthogonality of projected subspaces method \cite{yoon2006tops} were also proposed. Recently, some wideband DOA estimation methods based on sparse recovery have also been developed \cite{zhang2013wideband, wang2015novel, liu2011broadband, tang2011aliasing, gemba2017, antonello2019joint,nannuru2019sparse, gemba2019, wu2023gridless}. These sparsity-based methods have demonstrated superior performance compared to conventional methods and generally require much fewer samples. 

The multi-frequency model~\cite{tang2011aliasing,gemba2017, antonello2019joint,nannuru2019sparse, gemba2019, wu2023gridless, zhang2021enhanced} has shown success in modeling wideband signals. The multi-frequency model uses $N_f$ (rather than $1$) temporal frequency bins in a frequency set $\{F_1, \dots, F_{N_f} \}$ to characterize a wideband signal. All these frequencies are used for estimation, as opposed to using a single frequency under the narrowband model. One challenge for multi-frequency processing is aliasing \cite{nannuru2019sparse, wu2023gridless}, which will be present when the receiver spacing is greater than the half wavelength of the highest frequency. The performance of a DOA estimation method may degrade significantly in the presence of aliasing. In \cite{tang2011aliasing}, the authors present an aliasing-free DOA estimation method based on sparse signal recovery. In \cite{gemba2017, nannuru2019sparse, gemba2019}, wideband signal DOA estimation based on sparse Bayesian learning (SBL) with multiple frequencies is proposed \cite{nannuru2019sparse} and applied to matched field processing \cite{gemba2017} and robust ocean acoustic localization \cite{gemba2019}. \textcolor{black}{A DOA estimation method based on low-rank structured matrix completion for sparse arrays under the multi-frequency model is proposed in \cite{zhang2021enhanced}}. A joint localization and dereverberation method based on sparse regularization is also proposed in \cite{antonello2019joint} for room source localization and tracking.

\subsubsection{Atomic Norm Minimization (ANM)}
ANM was initially proposed in \cite{chandrasekaran2012convex} as a general framework for promoting sparse signal decompositions. The main benefit of ANM is that it overcomes the grid mismatch error that plagues grid-based methods. The pioneering ANM paper \cite{candes2014towards} proposed an optimization-based continuous  (temporal) frequency estimation method and provided a theoretical guarantee when full data are available. 
The authors in \cite{tang2013compressed} studied continuous temporal frequency estimation based on randomly sampled data for the single measurement vector (SMV) case. 
ANM for multiple measurement vectors (MMVs) under the uniform (or equispaced) time samples (analogous to a uniform linear array, or ULA) setup was studied in \cite{li2015off, yang2016exact,  yang2018sample}, and it was extended to the non-uniform array (NUA) setting in \cite{wagner2021}. It was also extended to multiple frequencies for wideband DOA estimation in \textcolor{black}{\cite{wu2023gridless, wu2022gridlessicassp, jiang2020gridless}. }
The sample complexity of modal analysis with random temporal compression was established in \cite{li2018atomic}. 
See \cite{chi2020harnessing} for a comprehensive overview of ANM and its applications. 

\subsubsection{Lifting}
\textcolor{black}{A ``lifting trick'' was applied to an ANM problem for the point spread function (PSF) estimation in \cite{chi2016guaranteed}. This allowed a bilinear inverse problem to be transformed into a linear inverse problem by assuming the PSF lies in a known low-dimensional space. This lifting idea was also applied to biconvex compressive sensing problems in \cite{ling2015self} so that these problems can be formulated as convex programs.}

\subsubsection{Non-uniform Array and More Sources than Sensors}
An NUA enables the possibility to resolve more sources than the number of physical sensors. Early works involving NUA include the minimum redundancy array (MRA) \cite{moffet1968minimum}, and the minimum holes array (MHA) \cite{bloom1977applications}. For a given number of sources, MRA and MHA require an extensive search through all possible sensor combinations to find the optimal design. Recently, a new structure of NUA, known as a co-prime array \cite{vaidyanathan2010sparse}, was developed. The co-prime array has a closed-form expression for the sensor positions so that the exhaustive search over the sensor combinations is avoided. The nested array \cite{pal2010nested} and co-array \cite{wang2016coarrays} based approaches were also proposed to detect more sources than the number of sensors.

An alternative way to resolve more sources than sensors is to use fourth-order cumulants \cite{dogan1995applications, chevalier2005virtual}. However, this approach is limited to non-Gaussian sources. In \cite{ma2009doa}, with the help of the Khatri-Rao (KR) product and assuming quasi-stationary sources, it was shown that one can identify up to $2N - 1$ sources using an $N$-element ULA without computing higher-order statistics. Unfortunately, the quasi-stationary assumption does not apply to stationary sources.

\subsection{Our Contributions}
In previous work \cite{wu2023gridless}, we developed a gridless DOA estimation method for the multi-frequency model based on ANM. This was formulated as a semi-definite program (SDP) problem so that ANM is solved using off-the-shelf SDP solvers, e.g. CVX \cite{grant2014cvx}. The DOAs are retrieved by finding the roots of the dual polynomial. The dual polynomial served as a certificate for the optimality and an interpolation method that constructed the dual certificate was presented.

In this work, we propose a wideband DOA estimation framework that significantly expands the applicability from ~\cite{wu2023gridless}.  Our contribution is summarized in the following respects (see also Table~\ref{tab:table1}).

\subsubsection{Regularization-free Framework} An SDP that is equivalent to ANM is formulated in \cite{wu2023gridless} based on the dual atomic norm and the definition of the dual polynomial in the noise-free case. When noise is present, a common strategy in ANM works is giving some tolerance to the constraints using regularization in the SDP \cite{xenaki2015, yang2016exact, tang2013compressed}. In addition to the challenges from the noise, \cite{wu2023gridless} shows that for array spacing above half a wavelength of the highest frequency and with multiple sources, the performance may degrade remarkably due to a phenomenon termed near collision \cite{wu2023gridless}. To mitigate near collisions, an $\ell_{1, 2}$ regularization term is added. Although these regularization terms prevent failures due to noise or near collisions, they lead to bias. The performance of ANM degrades due to such bias compared to the competing method SBL~\cite{gerstoft2016}, especially at a low signal-to-noise ratio (SNR). 

Although most ANM works promote robustness to noise by adding a regularization term, \cite{wagner2021} demonstrates that it is possible to deal with the noise by solving a noise-free optimization problem. In \cite{wagner2021}, the authors propose a two-step DOA estimation approach. The first step is to apply the alternating projection (AP) algorithm to solve a \textit{noise-free} optimization problem and obtain a matrix with an irregular Toeplitz structure. The second step computes an irregular Vandermonde decomposition (achieved by generalized root-MUSIC) to retrieve the DOAs from the irregular Toeplitz matrix. Although the optimization problem solved in the first step does not have explicit robustness to the noise, the second step enables the method to work in noisy cases. This method effectively avoids the explicit bias and the non-trivial effort required to tune the regularization parameter.

Inspired by \cite{wagner2021}, we formulate the dual problem of the \textit{noise-free} SDP in \cite[eq. (20)]{wu2023gridless} \textit{without regularization}. This problem is again an SDP and we deem it as the \textit{primal domain SDP} (since~\cite{wu2023gridless} formulates its SDP in the dual domain). \textcolor{black}{Both the primal and dual SDPs involve a certain lifting mapping which we define. Though different from~\cite{chi2016guaranteed}, this lifting embeds the problems in a higher dimensional space where it is more natural to combine all frequencies. } 

Solving this SDP gives a Toeplitz matrix, and the DOAs are further retrieved by Vandermonde decomposition of this Toeplitz matrix. One computational method for Vandermonde decomposition is root-MUSIC \cite{rao1989performance}, and it has robustness to both noise and near collisions. Therefore, with the help of ``post-SDP processing'' (root-MUSIC), regularization is avoided and no prior knowledge of the noise is needed. At low SNRs in simulation, the method can achieve a better performance than competing methods, and it approaches the Cram\'{e}r-Rao bound (CRB) \textcolor{black}{\cite{liang2020review, liang2021cramer} for Gaussian noise. }

\subsubsection{Non-uniform Frequencies and Irregular Vandermonde Decomposition} We also develop a fast SDP for the primal domain SDP. The fast SDP is derived based on the dual problem of the fast algorithm in \cite{wu2023gridless}. The fast algorithm can not only improve the speed but also can extend the method to the non-uniform frequency (NUF) case. In this case, the DOAs are encoded in a matrix with an \textit{irregular Toeplitz} structure. We apply the irregular Vandermonde decomposition (IVD)~\cite{wagner2021} to this matrix to retrieve the DOAs. Furthermore, we provide a theoretical guarantee for the existence of the IVD which is not shown in \cite{wagner2021}. While it is mentioned in \cite{wu2023gridless} that the fast dual algorithm proposed therein can be applied to the NUF case, this is not tested in~\cite{wu2023gridless}, and our experiments (see Fig.~\ref{RMSE_4plot}) indicate that the fast primal method is more effective.

\subsubsection{Multiple Snapshots and More Sources Than Sensors} The method in \cite{wu2023gridless} is developed under the SMV case. Prior works show that MMVs can give improved performance \cite{yang2016exact, yang2018sparse, li2015off}. That motivates us to extend the framework in \cite{wu2023gridless} to the MMV case. In the MMV setting, the received signal from the sensor array is a three-dimensional tensor (sensors $\times$ snapshots $\times$ frequencies). Based on the signal model, we formulate the corresponding ANM problem and derive the SDP (in the dual domain) that is equivalent to the ANM. The dual problem of the SDP is then derived to obtain the SDP in the primal domain. The purpose of the primal SDP is to enhance the robustness to the noise and near collision without regularization.

\begin{table*}[]
  \begin{center}    
    \caption{\label{tab:table1} Survey of SDPs used in \cite{wu2023gridless} and this work.}
    \begin{tabular}{lll} 
       \hline & \cite{wu2023gridless} & This work\\
      \hline\hline
      Assumption & ULA and uniform frequency & NUA and NUF\\
      \hline
      Procedure & Dual SDP $\rightarrow$ Polynomial rooting & Primal SDP $\rightarrow$  IVD \\
      \hline
      Model & SMV & MMV\\
      \hline
      Noise-free SDPs  &  Dual uniform (20) (SDP equivalent to ANM) &  Dual uniform \eqref{ssdp} (SDP equivalent to MMV-MF ANM) \\
      
                &  Fast dual (27)-(28) (extension of (20); nonuniform not tested)       &  Fast dual \eqref{ssdp_nua}, extension of\eqref{ssdp}; accommodates NUA/NUF   \\
                &   Full-dimension primal (29) (dual of (20); uniform case)   &  Fast primal \eqref{ssdp_fast_mmv} (dual of \eqref{ssdp_nua}; accommodates NUA/NUF)  \\
                &           & Full-dimension primal \eqref{ssdp_mmv} (dual of \eqref{ssdp}; uniform case) \\

      \hline
      Noisy SDPs & Dual uniform (21), robust version of (20)         & Dual uniform \eqref{ssdp_noise} (robust version of \eqref{ssdp})      
      \\
      \hline
    \end{tabular}
  \end{center}
\end{table*}

The multi-frequency setup also enables resolving more sources than sensors case in the ULA setting. The maximum number of uniquely identifiable sources in an $N_M$-element ULA is $N_M - 1$ \cite[Sec. 11.2.3]{yang2018sparse} for the single-frequency case. Co-prime array techniques \cite{vaidyanathan2010sparse} can break through such a limit with a carefully designed array structure, enabling the resolution of more sources than the number of sensors. We show that it is possible to resolve more sources than sensors \textit{with a ULA under the multi-frequency model}. The physical intuition is that multiple frequencies increase the diversity of the harmonics and these ``new harmonics'' can serve as extra ``virtual sensors'' in a large virtual array. Due to this intrinsic property, it is possible to break through such a bottleneck in the ULA setup. In many practical scenarios, the array geometry is fixed and ULA is one of the most commonly used arrays.  This result has a practical impact and demonstrates the benefit of multi-frequency processing. 

In summary, the framework proposed is superior to \cite{wu2023gridless} in terms of generality, practicality, performance, and complexity. Our work also demonstrates the possibility of resolving more sources than sensors \textit{under the ULA setup} which is an important merit of the multi-frequency model.

\subsection{Notation}
Throughout the paper, the following notation is adopted. Boldface letters are used to represent matrices and vectors. Conventional notations $(\cdot)^T$, $(\cdot)^H$, $(\cdot)^*$, $\langle \cdot, \cdot \rangle_{\mathbb{R}}$, and $\langle \cdot, \cdot \rangle$ stand for matrix/vector transpose, Hermitian transpose, complex conjugate, real inner product, and inner product, respectively. $\mathrm{Tr}(\cdot)$ is used to represent the trace of a matrix. $\| \cdot \|_p$, $\| \cdot \|_F$, and $\| \cdot \|_{\mathrm{HS}}$ are used to express vector $\ell_p$ norm, matrix Frobenius norm, and Hilbert-Schmidt norm for the tensor (for a 3D tensor $\|\mathcal{A} \|_{\mathrm{HS}} =\sqrt{\sum_{ijk}|a_{ijk}|^2}$). For a Hermitian matrix $\mathbf{A}$, $\mathbf{A} \succeq 0 $ means $\mathbf{A}$ is a positive semidefinite matrix. The imaginary unit is denoted by $j = \sqrt{-1}$. 

\section{Preliminaries}

When multiple snapshots are available, DOA estimation methods can have improved performance~\cite{li2015off, yang2016exact, yang2018sparse}. In this section, we extend the SMV multi-frequency ANM framework for gridless DOA estimation from~\cite{wu2023gridless} to the MMV setting; we refer to the resulting framework as the \textit{MMV-MF model}. This model will help us explore the possibility of having more sources than the sensors in Sec.~\ref{ula_source}.

\subsection{Assumptions}
\label{assump}
The following assumptions are made for the array configuration and signal model:
\begin{enumerate}
    \item The sensors comprise a linear array with positions drawn from a uniform grid ${\{0,1,\dots, N_M-1\}\cdot d}$, where $d$ is the sensor spacing unit.
    We let $\mathcal{M} \subseteq \{0,1, \dots, N_M-1\}$ denote the indices of the actual sensors; the resulting positions are thus $\{m \cdot d|m \in \mathcal{M}\}$. 
    We define $N_m := |\mathcal{M}| \le N_M$ as the number of sensors.
    When all sensors are present, $N_m = N_M$, and we have a {\em uniform linear array (ULA)} case. 
    When only some sensors are present, $N_m < N_M$, and we have a {\em nonuniform array (NUA)} case.
    \item The sources have temporal frequency components drawn from a uniform grid ${\{1, \dots, N_F\} \cdot F_1}$, where $F_1$ is the spacing between frequencies. 
    Let $\lambda_1 := c/F_1$ denote the wavelength corresponding to $F_1$, where $c$ is the propagation speed.
    We assume $\lambda_1 = 2d$ where $d$ is the sensor spacing unit above; equivalently, $d = \frac{c}{2F_1}$. 
    This spacing is for simplifying the derivation and can be relaxed to any $d \leq \frac{\lambda_1}{2}$ (see \cite{wu2023gridless} for details).
    We let $\mathcal{F} \subseteq \{1, \dots, N_F\}$ denote the indices of the active source frequencies; the resulting frequencies are thus $\{f \cdot F_1|f \in \mathcal{F}\}$ and the wavelengths are $\{\lambda_1/f|f \in \mathcal{F}\}$. 
    We define $N_f := |\mathcal{F}| \le N_F$ to be the number of active source frequencies. 
    When all frequencies are active, $N_f = N_F$, and we refer to this as the {\em uniform frequency} case. 
    When only some frequencies are active, $N_f < N_F$, and we refer to this as the {\em nonuniform frequency (NUF)} case.
    \item Suppose there are $N_l$ snapshots (time samples) received by each sensor. The source amplitude for the $f$-th frequency ($f \in \mathcal{F}$) is $\mathbf{x}_w(f) = [x_w^{(1)}(f) \;\dots\; x_w^{(N_l)}(f)]^T \in \mathbb{C}^{N_l}$.
    \item There are $K$ active uncorrelated sources impinging on the array from unknown directions of arrival (DOAs) $\theta$, or in directional cosines 
    \begin{equation} \label{eq:w}
     w := F_1 d \cos(\theta) / c = \cos(\theta) /2.
    \end{equation}
\end{enumerate}
\subsection{MMV-MF Model}

We begin by considering the case of a ULA with uniform frequencies, i.e.,  $N_m = N_M$ and $N_f=N_F$. (We incorporate the NUA and NUF cases in Section~\ref{fast}.)
The received signals can be arranged into a tensor $\mathcal{Y} \in \mathbb{C}^{N_M \times N_l \times N_F}$ (sensors $\times$ snapshots $\times$ frequencies) with the following structure: 
\begin{equation}
\label{y_model}
    \mathcal{Y} = \mathcal{X} + \mathcal{N}
\end{equation}
\begin{equation}
\begin{aligned}
    \mathcal{X} &= \sum_w c_w [\mathbf{a}(1, w)\mathbf{x}_w^T(1) | ... | \mathbf{a}(N_F, w) \mathbf{x}_w^T(N_F) ] \\
    &= \sum_w c_w \mathbf{A}(w) * \mathbf{X}_w^T
\end{aligned}
\label{eq:Xtensordef}
\end{equation}
where $\mathbf{a}(f, w) = [1 \; e^{-j2\pi w f} \dots e^{-j2\pi w f(N_M-1)}]^T = [1 \; z^f \dots z^{f(N_M - 1)}]^T \in \mathbb{C}^{N_M}$ ($z := e^{-j2 \pi w}$) is the array manifold vector for the $f$-th frequency. $\mathcal{N} \in \mathbb{C}^{N_M \times N_l \times N_F}$ denotes additive Gaussian uncorrelated noise in \eqref{y_model}. Denote $\mathbf{A}(w) = [\mathbf{a}(1, w) \dots \mathbf{a}(N_F, w)] \in \mathbb{C}^{N_M \times N_F}$ and $\mathbf{X}_w = [\mathbf{x}_w(1) \dots \mathbf{x}_w(N_F)]^T \in \mathbb{C}^{N_F \times N_l}$.  $\mathbf{A}(w) * \mathbf{X}_w^T$ is the ``reshaped Khatri-Rao product'' defined as $[\mathbf{A}(w) * \mathbf{X}_w^T]_{::f} := \mathbf{a}(f, w)\mathbf{x}_w^T(f)$ ($f = 1, ..., N_F$). When $N_l = 1$, the above matches the SMV model in~\cite{wu2023gridless}. \textcolor{black}{We assume $\| \mathbf{X}_w \|_F = 1$, as the coefficient $c_w$ can used to absorb any other scaling of the source amplitudes via the product $c_w \mathbf{X}_w$.}

Finally, we define
\begin{equation}
N = N_F (N_M-1) + 1,
\label{eq:Ndef}
\end{equation}
noting that $N_F (N_M-1)$ appears in the largest exponent of any array manifold vector used in the MMV-MF model. Consequently, $N$ will determine the size of certain SDP formulations such as~\eqref{ssdp}.

\subsection{Collision and Near Collision}

A challenge for multi-frequency processing is the risk of a phenomenon known as {\em collision}, which occurs when, at some frequencies, the array manifold vectors for two DOAs coincide due to aliasing. Two DOAs $w_1$ and $w_2$ are said to have a \textit{collision} in the $f$-th frequency if \cite[eq. (46)]{wu2023gridless}
\begin{equation}
    \mathbf{a}(f, w_1) = \mathbf{a}(f, w_2). 
\end{equation}
Such a collision occurs whenever $w_1$ and $w_2$ satisfy \cite[eq. (47)]{wu2023gridless}
\begin{equation}
        |w_1 - w_2| = \frac{k}{f} \quad (f \in \mathcal{F}, f > 1).
\end{equation}
A \textit{near collision} is said to occur when~\cite[eq. (50)]{wu2023gridless} 
\begin{equation}
    |w_1 - w_2|  \approx \frac{k}{f}  \quad (f \in \mathcal{F}, f > 1).
    \label{eq:nearcollisson}
\end{equation}

\subsection{Irregular Vandermonde and Toeplitz Matrices}
\label{Irregular}

Define some integer-valued vector $\bm{\gamma} = [\gamma_1 \dots \gamma_{N_\gamma}]^T \in \mathbb{Z}^{N_\gamma}$, complex-valued vector $\mathbf{z} = [z_1 \dots z_{N_z}]^T \in \mathbb{C}^{N_z}$, and $\mathbf{w}(\bm{\gamma}, z) := [z^{\gamma_1} \dots z^{\gamma_{N_\gamma}}]^T$. For arbitrary dimensions $N_\gamma$ and $N_z$, an {\em irregular Vandermonde matrix} of size $N_\gamma \times N_z$ is a matrix having the form~\cite[eq.\ (25)]{wagner2021}
\begin{align}
    \mathbf{W} = \mathbf{W}(\bm{\gamma}, \mathbf{z}) &= [\mathbf{z}^{\gamma_1} \dots \mathbf{z}^{\gamma_{N_\gamma}}]^T \nonumber \\ &= [\mathbf{w}(\bm{\gamma}, z_1) \dots \mathbf{w}(\bm{\gamma}, z_{N_z})].
\label{eq:ivdsetup}    
\end{align} 
Note that when the entries of $\bm{\gamma}$ form an arithmetic progression, specifically $\bm{\gamma} = [0 \dots N_\gamma - 1]^T$, $\mathbf{W}(\bm{\gamma}, \mathbf{z})$ forms a regular Vandermonde matrix.

An {\em $(N_\gamma,N_z)$-irregular Toeplitz matrix} is any matrix 
$\mathbf{T} \in \mathbb{C}^{N_\gamma \times N_\gamma}$ that can be constructed from an irregular Vandermonde matrix as follows~\cite[eq. (27)]{wagner2021}:
\begin{equation}
\label{IV}
    \mathbf{T} = \mathbf{W}(\bm{\gamma}, \mathbf{z})\mathbf{D} \mathbf{W}(\bm{\gamma}, \mathbf{z})^H, |\mathbf{z}| = 1,
\end{equation}
where $\bm{\gamma} \in \mathbb{Z}^{N_\gamma}$ and $\mathbf{z} \in \mathbb{C}^{N_z}$, and where $\mathbf{D} \in \mathbb{R}^{N_z \times N_z}$ is a diagonal matrix. We refer to~\eqref{IV} as an {\em irregular Vandermonde decomposition (IVD)}. Note that any $N_\gamma \times N_\gamma$ positive semi-definite regular Toeplitz matrix $\mathbf{T}$ with rank $N_z$ has a regular Vandermonde decomposition of the form~\eqref{IV} in which $\bm{\gamma} \in \mathbb{Z}^{N_\gamma}$ is an arithmetic progression.

\section{Atomic Norm Minimization for MMV-MF}
\label{anm_mf}
In this section, we formulate the atomic norm minimization problem for the MMV-MF model. Then, we derive an equivalent SDP that makes the proposed framework computationally feasible. \textcolor{black}{We note that the ANM we derive has multiple frequencies while the ANM in \cite{yang2016exact, li2015off} operates at a single frequency. The multi-frequency model can be applied to wideband signals, while their MMV model can only be applied to narrowband signals. }

Define the atomic set 
\begin{equation}
    \mathcal{A} = \{\mathbf{A}(w) * \mathbf{X}_w^T \;|\; w \in [-1/2, 1/2], \|\mathbf{X}_w\|_F = 1\}.
\end{equation}
The atomic norm of a tensor $\mathcal{X} \in \mathbb{C}^{N_M \times N_l \times N_F}$ is defined as  $\|\mathcal{X}\|_{\mathcal{A}} := \inf \{ \sum_w |c_w| \big| \mathcal{X} = c_w \mathbf{A}(w) * \mathbf{X}_w^T \;|\;  \|\mathbf{X}_w\|_F = 1 \}$.
The {\em atomic norm minimization (ANM)} problem for the noise-free case can be expressed as 
\begin{equation}
\label{anm}
    \min_{\mathcal{X}} \quad  \|\mathcal{X}\|_{\mathcal{A}} \quad  \textrm{s.t.}   \quad  \mathcal{Y} = \mathcal{X}.
\end{equation}
When noise is present, the optimization problem is modified to relax the equality constraint:
\begin{equation}
\label{anm_noise}
    \min_{\mathcal{X}} \quad  \|\mathcal{X}\|_{\mathcal{A}} \quad  \textrm{s.t.}   \quad   \|\mathcal{Y} - \mathcal{X}\|_{\mathrm{HS}} \leq \eta.
\end{equation}

The following proposition guarantees that \eqref{anm} is equivalent to an SDP problem. 

\begin{proposition}
\label{p1}
Problem \eqref{anm} is equivalent to the following SDP problem
\begin{equation}
\label{ssdp}
\begin{aligned}
    & \max_{\mathcal{Q}, \mathbf{P}_0} \langle \mathcal{Q}, \mathcal{Y} \rangle_{\mathbb{R}} 
    \quad \textrm{s.t.}  \left[                 
  \begin{array}{cc}   
    \mathbf{P}_0 & \widetilde{\mathbf{Q}} \\  
    \widetilde{\mathbf{Q}}^H & \mathbf{I}_{N_lN_F} \\  
  \end{array}
\right]  \succeq 0, \\
&\sum_{i = 1}^{N-k} \mathbf{P}_0(i, i+k) = \delta_k, \widetilde{\mathbf{Q}} = [\mathcal{R}(\mathbf{Q}_1) \dots  \mathcal{R}(\mathbf{Q}_{N_F})],
\end{aligned}
\end{equation}
\end{proposition}
where $\mathcal{Q} = [\mathbf{Q}_1 | \dots | \mathbf{Q}_{N_F}] \in \mathbb{C}^{N_M \times N_l \times N_F}$ is the dual variable, $\mathbf{P}_{0} \in \mathbb{C}^{N \times N}$, $\widetilde{\mathbf{Q}} = [\widetilde{\mathbf{Q}}_1 \dots \widetilde{\mathbf{Q}}_{N_F}] \in \mathbb{C}^{N \times N_l N_F}$, and $\widetilde{\mathbf{Q}}_f = \mathcal{R}(\mathbf{Q}_f): N_M \times N_l \rightarrow N \times N_l$ is a mapping defined as
\begin{equation}
\label{R_map}
   \! \mathcal{R}(\mathbf{Q}_f)(i, l) = \!\! \left\{
\begin{array}{ll}
 \!\!\!   \mathbf{Q}_f(m\!, l)  \!\!\!  &\mbox{for} \; (i, l)\!=\!(f(m\!- \!1) \! + \! 1, \! l)\\
\!\!\!    0  \!\!\! & \textrm{otherwise}.
\end{array}
\right.
\end{equation}
\\
\textit{Proof} See Appendix \ref{p_31}.

Fig. \ref{R_map_plot} demonstrates the mapping $\mathcal{R}$. Across all frequencies, $\mathcal{R}: N_M \times N_l \times N_F \rightarrow N \times N_lN_F$ is a linear mapping and can be expressed as a tall binary matrix multiply $\operatorname{vec}(\widetilde{\mathbf{Q}}) = \mathbf{R}\operatorname{vec}(\mathcal{Q})$. 
The transpose of the matrix $\mathbf{R}$ describes the behavior of the adjoint operator $\mathcal{R}^\ast:  N \times N_lN_F \rightarrow N_M \times N_l \times N_F$, which is also demonstrated in Fig.\ \ref{R_map_plot}.

To provide intuition for the role of $\mathcal{R}$, recall from~\eqref{eq:Xtensordef} that the array manifold vectors in the MMV-MF model are frequency-dependent, and so the rows of different slices of $\mathcal{X}$ correspond to different space-frequency products. After lifting the dual variable tensor $\mathcal{Q}$ to a higher-dimensional space, however, every row corresponds to the same space-frequency product $f(m-1)$, allowing $\widetilde{\mathbf{Q}}$ to play a similar role in the SDP to the dual variables in more conventional ANM formulations.

In the noisy case, the equivalent SDP of \eqref{anm_noise} is the regularized version of \eqref{ssdp}:
\begin{equation}
\label{ssdp_noise}
\begin{aligned}
    & \max_{\mathcal{Q}, \mathbf{P}_0}  \langle \mathcal{Q}, \mathcal{Y} \rangle_{\mathbb{R}} - \eta \| \mathcal{Q} \|_{\mathrm{HS}}
    \quad \textrm{s.t.}  \left[                 
  \begin{array}{cc}   
    \mathbf{P}_0 & \widetilde{\mathbf{Q}} \\  
    \widetilde{\mathbf{Q}}^H & \mathbf{I}_{N_lN_F} \\  
  \end{array}
\right]  \succeq 0, \\
&\sum_{i = 1}^{N-k} \mathbf{P}_0(i, i+k) = \delta_k, \widetilde{\mathbf{Q}} = [\mathcal{R}(\mathbf{Q}_1) \dots  \mathcal{R}(\mathbf{Q}_{N_F})],
\end{aligned}
\end{equation}
where $\eta$ depends on the noise level and is the same as in \eqref{anm_noise}.
 
\section{Regularization-free SDP and Fast Algorithm}
\label{fast}
In the previous section, we obtained an SDP that is equivalent to ANM. This SDP relies on the dual norm and dual polynomial (see Appendix \ref{p_31} for the dual norm \eqref{x_anm} and dual polynomial \eqref{x_poly}), and so we deem the SDP in \eqref{ssdp} as the \textit{dual SDP}. We now derive the dual problem of the SDP in Sec. \ref{anm_mf}; we deem this as the \textit{primal SDP}. The benefit of the primal SDP is that it is regularization-free and it thus avoids regularization bias in \eqref{ssdp_noise}. In numerical experiments, this primal SDP is inherently robust to noise and near collisions. Further, we derive a fast, reduced-dimension version of the primal SDP. The fast program improves the speed, and more importantly, it relaxes the requirements that the sensor positions and temporal frequencies be uniform.

\subsection{Non-uniform Array (NUA) and Non-uniform Frequency (NUF) Settings}
\label{nua_nuf}
In the previous sections, we focused on the ULA and uniform frequency case. However, in general, the array spacing and frequency may not be uniform. Thus we generalize the proposed framework to NUA and NUF cases. 

Recall that $\mathcal{F} \subseteq \{1, \dots, N_F\}$ denotes the indices of the active source frequencies, with $N_f := |\mathcal{F}| \le N_F$ denoting the number of active frequencies. The nonuniform frequency (NUF) case corresponds to the scenario where $N_f < N_F$, i.e., only some of the frequencies are active. Similarly, $\mathcal{M} \subseteq \{0,1, \dots, N_M-1\}$ denotes the indices of the sensors, with $N_m := |\mathcal{M}| \le N_M$ denoting the number of sensors. The nonuniform array (NUA) case corresponds to the scenario where $N_m < N_M$, i.e., only some sensors are present.

Recall that every exponent in an array manifold vector from the MMV-MF model involves a product of one temporal frequency and one sensor position. 
To capture all such products in the nonuniform setting, we define a spatial-frequency index set $\mathcal{U}$ as follows:
\begin{equation}
\mathcal{U} = \{m \cdot f |m \in \mathcal{M}, f \in \mathcal{F} \}.
\end{equation}
The cardinality of this set $N_u := |\mathcal{U}| \le N$, with $N$ is defined in~\eqref{eq:Ndef}. In many settings, $N_u \ll N$. In later sections, we see that the size of the fast SDP depends on $N_u$, and its complexity is greatly reduced compared to the original SDP.

\subsection{Fast Dual SDP for the NUA and NUF Case}
Proposition \ref{p1} gives the SDP for the ULA and uniform frequency case. We generalize the SDP to the NUA and NUF cases. The SDP is not only more general but also can reduce the complexity in the ULA and uniform frequency case. Inspired by the fast algorithm in \cite[Sec. III-F]{wu2023gridless}, the SDP in this section is considered the fast algorithm for MMV.

\begin{figure}[!t]
\centering
\includegraphics[width=8.5cm]{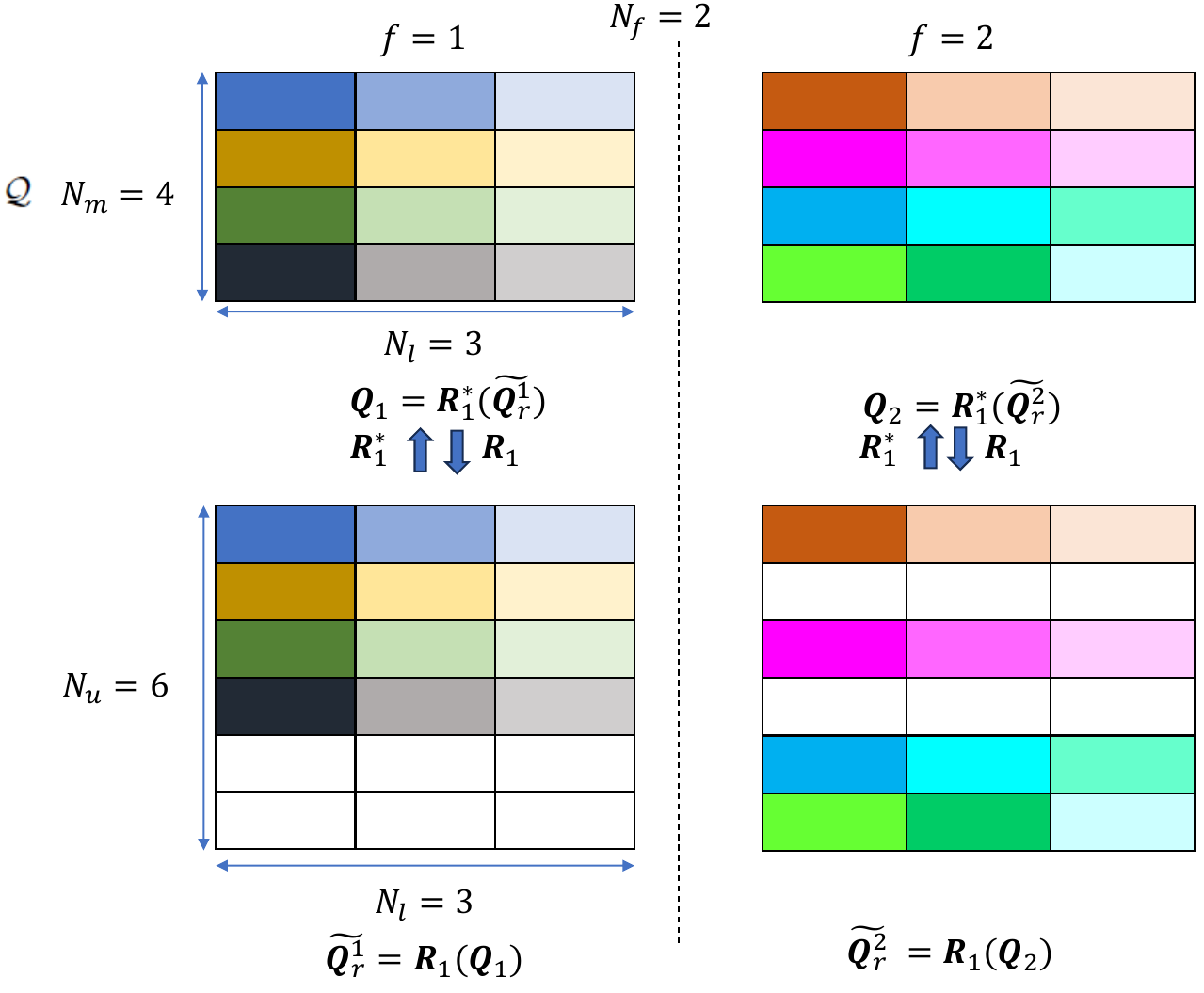}
\caption{Demonstration for the $\mathcal{R}_1(\cdot)$ mapping and its adjoint mapping $\mathcal{R}_1^*(\cdot)$. $N_m = N_M = 4$, $N_l = 3$, $N_f = N_F = 2$, $\mathcal{U} = \{0, 1, 2, 3, 4, 6 \}$, $N_u = |\mathcal{U}| = 6$. }
\label{R1_map}
\end{figure}

For NUA and NUF, the measurement tensor $\mathcal{Y} \in \mathbb{C}^{N_m \times N_l \times N_f}$ and the SDP in Proposition \ref{p1} is generalized as 
\begin{equation}
\label{ssdp_nua}
\begin{aligned}
    & \max_{\mathcal{Q}, \mathbf{P}_{r0}} \langle \mathcal{Q}, \mathcal{Y} \rangle_{\mathbb{R}} 
    \quad \textrm{s.t.}  \left[                 
  \begin{array}{cc}   
    \mathbf{P}_{r0} & \widetilde{\mathbf{Q}}_r \\  
    \widetilde{\mathbf{Q}}_r^H & \mathbf{I}_{N_lN_f} \\  
  \end{array}
\right]  \succeq 0, \\
&\sum_{\mathcal{U}_j - \mathcal{U}_i = k} \mathbf{P}_{r0}(i, j) = \delta_k, \widetilde{\mathbf{Q}}_r = [\mathcal{R}_1(\mathbf{Q}_1) \dots  \mathcal{R}_1(\mathbf{Q}_{N_f})],
\end{aligned}
\end{equation}
where $\mathcal{Q} = [\mathbf{Q}_1 | \dots | \mathbf{Q}_{N_f}] \in \mathbb{C}^{N_m \times N_l \times N_f}$ is the dual variable,  $\mathbf{P}_{r0} \in \mathbb{C}^{N_u \times N_u}$, $\widetilde{\mathbf{Q}}_r = [\widetilde{\mathbf{Q}}_r^1 \dots \widetilde{\mathbf{Q}}_r^{N_f}] \in \mathbb{C}^{N_u \times N_l N_f}$ ($\widetilde{\mathbf{Q}}_r^f = \mathcal{R}_1(\mathbf{Q}_f) \in \mathbb{C}^{N_u \times N_l}$), and $\mathcal{R}_1(\mathbf{Q}_f): N_m \times N_l \rightarrow N_u \times N_l$ is a mapping that pads zeros to the extra entries defined as
\begin{equation}
\label{R_1}
    \mathcal{R}_1(\mathbf{Q}_f)(r, l) =\!\! \left\{
\begin{array}{ll}
    \mathbf{Q}_f(m\!, l)  \!\!  &\mbox{for} \; (\mathcal{U}_r,l)\!=\!(f \cdot (m\!- \!1),  l)\\
    0  \!\! & \textrm{otherwise}.
\end{array}
\right.
\end{equation}
Fig. \ref{R1_map} demonstrates the $\mathcal{R}_1(\cdot)$ mapping. 
We note that any rows of $\widetilde{\mathbf{Q}}_f$ which would have remained all-zero under the operator $\mathcal{R}(\cdot)$ (corresponding to unused space-frequency products) are simply omitted in $\mathcal{R}_1(\cdot)$.
\begin{figure}[!t]
\centering
\includegraphics[width=8.5cm]{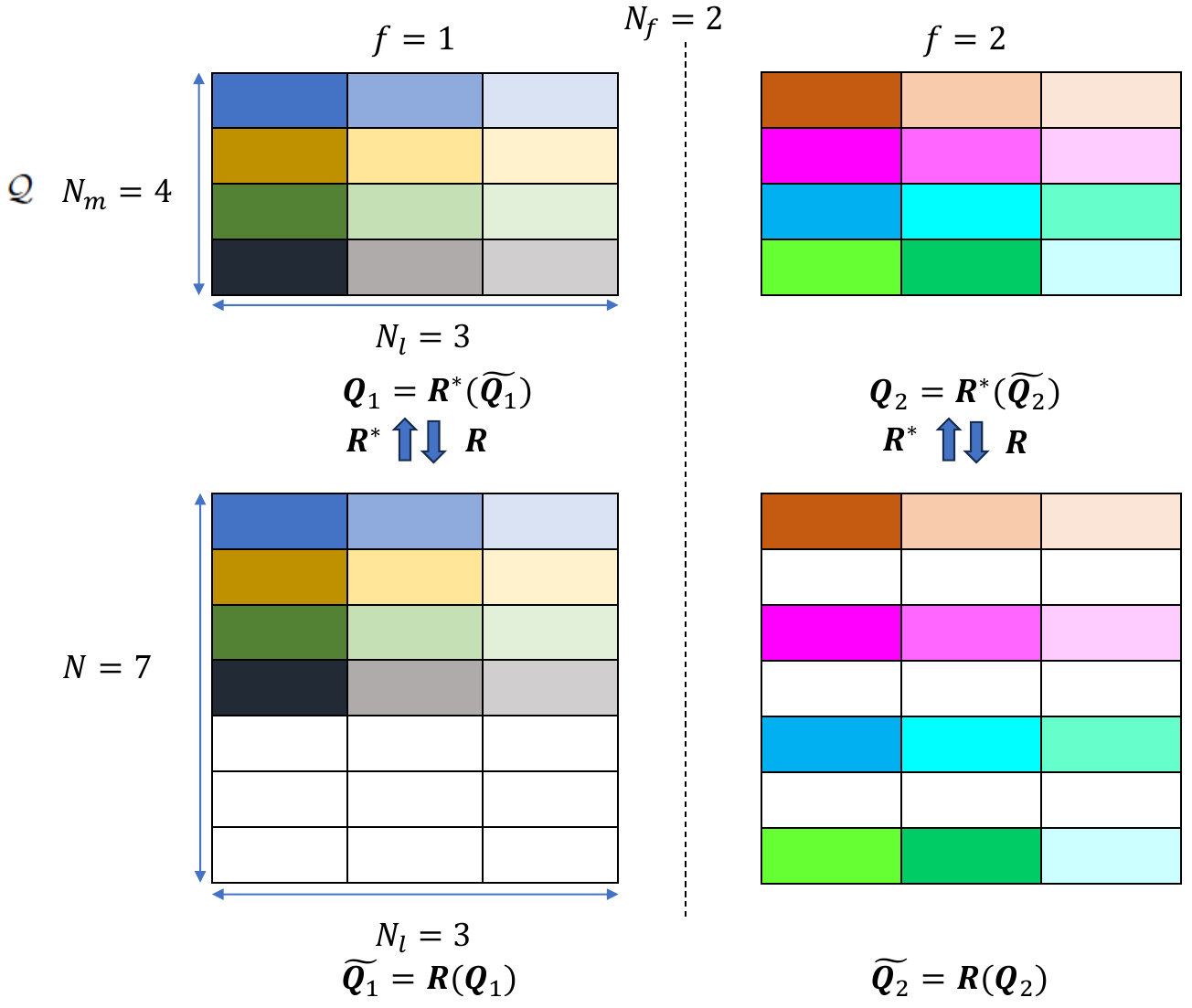}
\caption{Demonstration for the \textcolor{black}{lifting} mapping $\mathcal{R}(\cdot)$ and its adjoint mapping $\mathcal{R}^*(\cdot)$. $N_m = N_M = 4$, $N_l = 3$, $N_f = N_F = 2$, $N = (N_M - 1)N_F + 1 = 7.$}
\label{R_map_plot}
\end{figure}

Comparing \eqref{R_1} with \eqref{R_map}, these two mappings pad zeros for the same input $\mathbf{Q}_f$ to obtain the output matrix with a different dimension. The $\mathcal{R}$ mapping defined in \eqref{R_map} maps a matrix with $N_M$ rows into one with $N$ rows, while $\mathcal{R}_1$ defined in~\eqref{R_1} omits the unused products of temporal frequency and sensor position, mapping a matrix with $N_m$ rows into one with only $N_u$ rows. This not only gives a lower-dimensional formulation (the size of $\mathbf{P}_{r0}$ decreases from $N \times N$ to $N_u \times N_u$), but it naturally accommodates the NUA and NUF settings. Still, \eqref{ssdp_nua} can be applied to the ULA and uniform frequency case, where $N_u$ will often be somewhat smaller than $N$.

\subsection{Fast Primal SDP for the NUA and NUF Case}
\label{dual_sdp}

In this section, we derive the dual problem of \eqref{ssdp_nua}, yielding a fast primal SDP that is regularization-free and naturally accommodates the NUA and NUF settings. 

\begin{proposition}
\label{fastprimalSDP}
The dual problem of~\eqref{ssdp_nua} is given by \begin{equation}
\label{ssdp_fast_mmv}
\begin{aligned}
     &\min_{\mathbf{W}, \mathbf{u}, \mathbf{\widetilde{Y}}} [\mathrm{Tr(T}(\mathbf{u})) + \mathrm{Tr}(\mathbf{W})  ] \\
    \quad &\textrm{s.t.} 
\left[                 
  \begin{array}{cc}   
    \mathrm{T}(\mathbf{u}) & \mathbf{\widetilde{Y}} \\  
    \mathbf{\widetilde{Y}}^H &  \mathbf{W}\\  
  \end{array}
\right]  \succeq 0, \mathbf{Y}_f = \mathcal{R}_1^*(\mathbf{\widetilde{Y}}_f), f \in \mathcal{F},
\end{aligned}
\end{equation}
where $\mathbf{\widetilde{Y}} \in \mathbb{C}^{N_u \times N_lN_f}$, $\mathbf{W} \in \mathbb{C}^{N_lN_f \times N_lN_f}$, $\mathbf{Y}_f \in \mathbb{C}^{N_m \times N_l}$ is the slice of the received signal tensor $\mathcal{Y}$ corresponding to frequency $f$, and $\mathbf{\widetilde{Y}}_f \in \mathbb{C}^{N_u \times N_l}$ comes from taking the $N_l$ columns of $\mathbf{\widetilde{Y}}$ corresponding to frequency $f$. $\mathcal{R}_1^*(\cdot): N_u \times N_l \rightarrow N_m \times N_l$ is the adjoint mapping of $\mathcal{R}_1$.

\end{proposition}
\textit{Proof}~Consider the Lagrangian given by
\begin{equation}
\label{L_fun}
\begin{aligned}
     &\mathcal{L}(\mathcal{Q}, \mathbf{P}_{r0}, \mathbf{U}_r, \mathbf{\Lambda}_1, \mathbf{\Lambda}_2, \mathbf{\Lambda}_3,  \mathbf{\Lambda}_Q, \mathbf{v}) = \\
     & \langle \mathcal{Q}, \mathcal{Y} \rangle_{\mathbb{R}} - \bigg \langle \left[  
  \begin{array}{cc}   
    \mathbf{\Lambda}_1 & \mathbf{\Lambda}_2\\  
    \mathbf{\Lambda}_2^H & \mathbf{\Lambda}_3 \\  
  \end{array}
\right], \left[                 
  \begin{array}{cc}   
    \mathbf{P}_{r0} & \mathbf{U}_r \\  
    \mathbf{U}_r^H & \mathbf{I}_{N_lN_f} \\  
  \end{array}
\right] \bigg \rangle_{\mathbb{R}} \\
     &- \sum_{k = 0}^{N-1}v_k (\delta_k \!\!-\!\! \sum_{\substack{\mathcal{U}_j - \mathcal{U}_i = k}}\mathbf{P}_{r0}(i, j)) \!- \! \sum_{f \in \mathcal{F}} \langle \mathbf{\Lambda}_Q^f, \mathbf{U}_r^f \!- \! \mathcal{R}_1(\mathbf{Q}^f) \rangle_{\mathbb{R}} \\
     &= \sum_{f \in \mathcal{F}} [\langle \mathbf{Q}_f, \! \mathbf{Y}_f \rangle_{\mathbb{R}} \! + \! \langle \mathbf{\Lambda}_Q^f, \mathcal{R}_1(\mathbf{Q}_f) \rangle_{\mathbb{R}} ]\!\! - \!\! [\langle \mathbf{P}_{r0}, \mathbf{\Lambda}_1  \rangle_{\mathbb{R}} \!  \\ 
     &+ \! 2 \langle \mathbf{\Lambda}_2, \!\mathbf{U}_r \rangle_{\mathbb{R}} \!+\! \mathrm{Tr}(\mathbf{\Lambda}_3)]  \!-\!\! \mathbf{v}_0 \!+ \! \langle \mathbf{P}_{r0} , \! \mathrm{T}(\mathbf{v}) \rangle_{\mathbb{R}} \!-\!\! \sum_{f \in \mathcal{F}} \langle \mathbf{\Lambda}_Q^f, \! \mathbf{U}_r^f \rangle_{\mathbb{R}}.
\end{aligned}
\end{equation}
Note that we use the following fact during the derivation: $\sum_{k = 0}^{N-1}v_k \sum_{\mathcal{U}_j - \mathcal{U}_i = k}\mathbf{P}_{r0}(i, j) = \langle \mathbf{P}_{r0}, \mathrm{T}(\mathbf{v}) \rangle_{\mathbb{R}}$, where $\mathrm{T}: N \times 1 \rightarrow N_u \times N_u$  is explicitly defined as (note $^*$ denotes complex conjugate)
\begin{equation}
\label{T_map}
    \mathrm{T}(\mathbf{v})(i, j) :=  \left\{
\begin{array}{ll}
     v_{\mathcal{U}_j - \mathcal{U}_i }   \quad   & \;\; \mathcal{U}_j - \mathcal{U}_i \geq 0 \\
        v^*_{\mathcal{U}_i - \mathcal{U}_j } &\mathcal{U}_j - \mathcal{U}_i < 0.
\end{array}
\right.
\end{equation}
We provide an example of this mapping in Sec. \ref{T_eg}.

The dual matrix $\left[                 
  \begin{array}{cc}   
    \mathbf{\Lambda}_1 & \mathbf{\Lambda}_2\\  
    \mathbf{\Lambda}_2^H & \mathbf{\Lambda}_3 \\  
  \end{array}
\right]$ associated with the inequality constraint $\left[                 
  \begin{array}{cc}   
    \mathbf{P}_{r0} & \mathbf{U}_r \\  
    \mathbf{U}_r^H & \mathbf{I}_{N_f} \\  
  \end{array}
\right]  \succeq 0$ needs to be a PSD matrix to ensure that the inner product between these two matrices is non-negative so that the optimal value for the dual problem gives a lower bound for the primal problem.

The dual function is 
\begin{equation}
\label{dual_fun}
\begin{aligned}
 g(\mathbf{\Lambda}_1, \! \mathbf{\Lambda}_2, \! \mathbf{\Lambda}_3, \!\mathbf{\Lambda}_Q, \! \mathbf{v}) &= \!\!\! \inf_{\mathcal{Q}, \! \mathbf{P}_{r0}, \mathbf{U}_r} \!\!\!\! \mathcal{L}(\mathcal{Q}, \mathbf{P}_{r0},\! \mathbf{U}_r,\! \mathbf{\Lambda}_1,\! \mathbf{\Lambda}_2,\! \mathbf{\Lambda}_3,\! \mathbf{\Lambda}_Q, \! \mathbf{v}) \\
 & \mathrm{s.t.} \left[                 
  \begin{array}{cc}   
    \mathbf{\Lambda}_1 & \mathbf{\Lambda}_2\\  
    \mathbf{\Lambda}_2^H & \mathbf{\Lambda}_3 \\  
  \end{array}
\right] \succeq 0.
\end{aligned}
\end{equation}

The infimum of $\mathcal{L}$ in \eqref{L_fun} over $\mathcal{Q}$ is thereby $\inf_{\mathcal{Q}} J(\mathcal{Q}) :=  \sum_{f \in \mathcal{F}} [\langle \mathbf{Q}_f, \mathbf{Y}_f \rangle_{\mathbb{R}} + \langle \mathbf{\Lambda}_Q^f, \mathcal{R}_1^*(\mathbf{Q}_f) \rangle_{\mathbb{R}} ] =  \sum_{f \in \mathcal{F}} [\langle \mathbf{Q}_f, \mathbf{Y}_f \rangle_{\mathbb{R}} + \langle \mathbf{Q}_f, \mathcal{R}_1^*(\mathbf{\Lambda}_Q^f) \rangle_{\mathbb{R}} ] = \sum_{f \in \mathcal{F}} \langle \mathbf{Q}_f,  \mathbf{Y}_f + \mathcal{R}_1^*(\mathbf{\Lambda}_Q^f)  \rangle_{\mathbb{R}}$. The infimum of $J(\mathbf{Q})$ is bounded only if $\mathbf{Y}_f = -\mathcal{R}_1^*(\mathbf{\Lambda}_Q^f)$ for any $f \in \mathcal{F}$. Similarly, the infimum of $\mathcal{L}$ over $\mathbf{P}_{r0}$ is bounded only if $\mathrm{T}(\mathbf{v}) =  \mathbf{\Lambda}_1 \succeq 0.$ The infimum of $\mathcal{L}$ over $\mathbf{U}_r$ is bounded only if $\mathbf{\Lambda}_Q^f = -2\mathbf{\Lambda}_2^f$. Considering $2\mathbf{\Lambda}_2^f = \widetilde{\mathbf{Y}}_f$, then we must have $\mathbf{Y}_f = -\mathcal{R}_1^*(\mathbf{\Lambda}_Q^f) = \mathcal{R}_1^*(2\mathbf{\Lambda}_2^f) = \mathcal{R}_1^*(\widetilde{\mathbf{Y}}_f)$.

Considering $\mathbf{\Lambda}_3 = \frac{1}{2}\mathbf{W}$ and $\mathbf{v} = \frac{1}{2}\mathbf{u}$,  the dual function becomes $-\frac{1}{2}\mathrm{Tr}(\mathbf{W}) -\frac{1}{2}\mathrm{Tr}(\mathrm{T}(\mathbf{u})).$

Therefore, the fast program in the primal domain is given by~\eqref{ssdp_fast_mmv}. $\hfill\square$ \\

\subsubsection{SMV Setup} The fast program \eqref{ssdp_fast_mmv} can not only improve the execution time in the uniform cases, but it naturally accommodates the NUA and NUF cases as well. \eqref{ssdp_fast_mmv} can also be adapted to the SMV setup (i.e. $N_l = 1$). In that case, the received signal $\mathbf{Y}$ will reduce to an $N_m \times N_f$ matrix and \eqref{ssdp_fast_mmv} will reduce to 
\begin{equation}
\label{ssdp_fast_smv}
\begin{aligned}
     &\min_{\mathbf{W}, \mathbf{u}, \mathbf{\widetilde{Y}}} [\mathrm{Tr(T}(\mathbf{u})) + \mathrm{Tr}(\mathbf{W}) ] \\
    \quad &\textrm{s.t.} 
\left[                 
  \begin{array}{cc}   
    \mathrm{T}(\mathbf{u}) & \mathbf{\widetilde{Y}} \\  
    \mathbf{\widetilde{Y}}^H &  \mathbf{W}\\  
  \end{array}
\right]  \succeq 0, \mathbf{Y}_f = \mathcal{R}_1^*(\mathbf{\widetilde{Y}}_f), f \in \mathcal{F},
\end{aligned}
\end{equation} \\
where $\mathbf{\widetilde{Y}} \in \mathbb{C}^{N_u \times N_f}$, $\mathbf{W} \in \mathbb{C}^{N_f \times N_f}$, $\mathbf{Y}_f \in \mathbb{C}^{N_m \times 1}$ is the column of the received signal $\mathbf{Y}$ corresponding to frequency $f$, and $\mathbf{\widetilde{Y}}_f \in \mathbb{C}^{N_u \times 1}$ comes from taking the column of $\mathbf{\widetilde{Y}}$ corresponding to frequency $f$. 

\subsubsection{Comparison to full-dimension primal SDP} 
Recall that~\eqref{ssdp} is the dual SDP for the ULA and uniform frequency setting. A significant difference between \eqref{ssdp} and \eqref{ssdp_nua} lies in the dimensions of the matrices in the PSD constraint and the equality constraint. Following the same procedure in this section, the dual SDP of \eqref{ssdp} can be obtained, yielding the following full-dimension primal SDP for the ULA and uniform frequency case: 
\begin{equation}
\label{ssdp_mmv}
\begin{aligned}
     &\min_{\mathbf{W}, \mathbf{u}, \mathbf{\widetilde{Y}}_N} [\mathrm{Tr(Toep}(\mathbf{u})) + \mathrm{Tr}(\mathbf{W})] \\
    \quad &\textrm{s.t.} 
\left[                 
  \begin{array}{cc}   
    \!\!\!\!\!\mathrm{Toep}(\mathbf{u}) &   \!\!\!\!\!\mathbf{\widetilde{Y}}_N \\  
     \mathbf{\widetilde{Y}}^H_N &   \!\!\!\!\!\mathbf{W}\\  
  \end{array}
\right]  \succeq 0, \mathbf{Y}_f = \mathcal{R}^*(\mathbf{\widetilde{Y}}_{Nf}), f = 1, \dots, N_F,
\end{aligned}
\end{equation}
where $\mathrm{Toep}(\cdot): N \times 1 \rightarrow N \times N$ is the Toeplitz operator that maps a vector to a self-adjoint Toeplitz matrix. $\mathbf{\widetilde{Y}}_N \in \mathbb{C}^{N \times N_l N_F}$, $\mathcal{R}^*(\cdot): N \times N_l \rightarrow N_M \times N_l$ is the adjoint mapping of $\mathcal{R}(\cdot)$, and $\mathbf{\widetilde{Y}}_{Nf} \in \mathbb{C}^{N \times N_l}$ is taking $N_l$ columns from $\mathbf{\widetilde{Y}}_N$ (from the $(f - 1)\cdot N_l + 1$-th to the $f \cdot N_l$-th column).
Compared to \eqref{ssdp_fast_mmv}, a main  difference is that $\mathrm{T}(\mathbf{u})\in \mathbb{C}^{N_u\times N_u}$ in \eqref{ssdp_fast_mmv} is changed to $\mathrm{Toep}(\mathbf{u})\in \mathbb{C}^{N \times N}$. 

\subsection{Existence of Irregular Vandermonde Decomposition (IVD)}
\label{IVD_exist}

The full-dimension primal SDP in~\eqref{ssdp_mmv} has an interesting connection to the SDPs from the ANM literature which involve trace minimization of a (regular) Toeplitz matrix~\cite{tang2013compressed, yang2016exact}. In ANM problems that involve trace minimization of a regular Toeplitz matrix, one typically computes the Vandermonde decomposition of the resulting Toeplitz matrix in order to extract the frequencies/DOAs. Indeed, as we discuss further in Section~\ref{sec:rankdiscussion}, trace minimization serves as a convex relaxation of rank minimization, and a formal connection can be established between rank minimization and finding the sparsest decomposition in the atomic set~$\mathcal{A}$. 

In contrast, the fast primal SDPs~\eqref{ssdp_fast_mmv} and~\eqref{ssdp_fast_smv} derived in the previous section involve trace minimization not of a Toeplitz matrix but rather a matrix of the form $\mathrm{T}(\mathbf{u})$. (See Sec.~\ref{T_eg} for an illustration of the structure of $\mathrm{T}(\mathbf{u})$.) However, as we establish in Theorem~\ref{IVD_theo} below, there is an important connection between $\mathrm{T}(\mathbf{u})$ and Toeplitz matrices:  $\mathrm{T}(\mathbf{u})$ is guaranteed to be an irregular Toeplitz matrix, and therefore is guaranteed to have an IVD. This inspires our proposed method for extracting DOA information from $\mathrm{T}(\mathbf{u})$, which we outline in Section~\ref{sec:doa}.

\begin{theorem}
\label{IVD_theo}
For any $\mathbf{u}$ such that $\mathrm{Toep}(\mathbf{u})$ is PSD, $\mathrm{T}(\mathbf{u}) \in \mathbb{C}^{N_u \times N_u}$ is an $(N_u,K)$-irregular Toeplitz matrix, where $K = \mathrm{rank}(\mathrm{Toep}(\mathbf{u}))$. Specifically, $\mathrm{T}(\mathbf{u})$ has an IVD of the form~\eqref{IV}, where $\bm{\gamma} = [\mathcal{U}_1, \dots \mathcal{U}_{N_u}]^T$.

\end{theorem}

\textit{Proof} First, let $P_\mathcal{U} : \mathbb{C}^N \rightarrow \mathbb{C}^{N_u}$ denote a linear restriction operator that selects only the entries in a vector corresponding to the positions indexed by $\mathcal{U}$.

Now, consider $\mathrm{Toep}(\mathbf{u}) \in \mathbb{C}^{N \times N}$, and observe that $\mathrm{T}(\mathbf{u})\in \mathbb{C}^{N_u \times N_u}$ can be obtained by a mapping from $\mathrm{Toep}(\mathbf{u})$ as follows: $\mathrm{T}(\mathbf{u}) := P_\mathcal{U} \mathrm{Toep}(\mathbf{u}) P_\mathcal{U}^H$. Since $\mathrm{Toep}(\mathbf{u})$ is PSD, it is guaranteed to have a Vandermonde decomposition of the form~\cite[Theorem 11.5]{yang2018sparse}:
\begin{equation}
\label{vander}
\mathrm{Toep}(\mathbf{u}) = \mathbf{V}(\mathbf{z})\mathbf{DV}(\mathbf{z})^H
\end{equation}
where $\mathbf{V}(\mathbf{z}) \in \mathbb{C}^{N \times K}$ is a Vandermonde matrix parameterized by $\mathbf{z}$ with $|\mathbf{z}| = 1$, and $\mathbf{D} \in \mathbb{R}^{K \times K}$ is a diagonal matrix with positive diagonals. Hence,
\begin{equation}
\begin{aligned}
     &\mathrm{T}(\mathbf{u}) = P_\mathcal{U} \mathrm{Toep}(\mathbf{u}) P_\mathcal{U}^H = P_\mathcal{U} \mathbf{V}(\mathbf{z})\mathbf{DV}(\mathbf{z})^H P_\mathcal{U}^H  \\
     &= (P_\mathcal{U} \mathbf{V}(\mathbf{z}))\mathbf{D}(\mathbf{V}(\mathbf{z})^H P_\mathcal{U}^H) \\
     &=\mathbf{W}(\bm{\gamma}, \mathbf{z}) \mathbf{D} \mathbf{W}(\bm{\gamma}, \mathbf{z})^H,
\end{aligned}
\end{equation}
where $\mathbf{W}(\bm{\gamma}, \mathbf{z}) := 
P_\mathcal{U} \mathbf{V}(\mathbf{z})$ will be an irregular Vandermonde matrix of the form~\eqref{eq:ivdsetup} with $\bm{\gamma} = [\mathcal{U}_1 \dots \mathcal{U}_{N_u}]^T$. Therefore, $\mathrm{T}(\mathbf{u})$ is an $(N_u,K)$-irregular Toeplitz matrix. $\hfill\square$

\subsection{An Example for $\mathrm{T}(\mathbf{v})$} 
\label{T_eg}
We demonstrate the structure of $\mathrm{T}(\mathbf{v})$ in the following example. Consider $\mathcal{M} = \{0, 1, 3, 4\}$ and $\mathcal{F} = \{1, 3, 4 \}$. 
Therefore, $N_m = |\mathcal{M}| = 4$, $N_f = |\mathcal{F}| = 3$, $\mathcal{U} = \{0, 1, 3, 4, 9, 12, 16\}$, $N_M = 5$, $N_F = 4$, $N = (N_M - 1)N_F + 1  = 17$, and $N_u = |\mathcal{U}| = 7$. For $\mathbf{v} = [v_0 \dots v_{16}]^T$, $\mathrm{T}(\mathbf{v}) \in \mathbb{C}^{N_u \times N_u}$ can be expressed as 
$$ \mathrm{T}(\mathbf{v}) = 
\begin{bmatrix}
    v_0 & v_1 & v_3 & v_4 & v_{9} & v_{12} & v_{16} \\
    v_1^* & v_0 & v_2 & v_3 & v_8 & v_{11} & v_{15} \\
    v_3^* & v_2^* & v_0 & v_1 & v_6 & v_{9} & v_{13} \\
    v_4^* & v_3^* &  v_1^*& v_0 & v_5 & v_8 & v_{12} \\
    v_{9}^* & v_8^* & v_6^* & v_5^*& v_0 & v_3 & v_7\\
    v_{12}^* & v_{11}^* & v_{9}^* & v_8^* & v_3^* & v_0 & v_4 \\
    v_{16}^* & v_{15}^* & v_{13}^* & v_{12}^* & v_7^* & v_4^* & v_0 
\end{bmatrix}.
$$
Note $v_{10}$ and $v_{14}$ do not appear in $\mathrm{T}(\mathbf{v})$. In Theorem~\ref{IVD_theo}, we show that for any $\mathbf{v}$ such that such that $\mathrm{Toep}(\mathbf{v})$ is PSD, $\mathrm{T}(\mathbf{v})$ is guaranteed to be an irregular Toeplitz matrix. In this case, $\mathrm{T}(\mathbf{v})$ is guaranteed to have an IVD: $\mathrm{T}(\mathbf{v}) = \mathbf{W}(\bm{\gamma}, \mathbf{z}) \mathbf{D} \mathbf{W}(\bm{\gamma}, \mathbf{z})^H$, where $\mathbf{D} = \mathrm{diag}(d_1, \dots, d_K)$ with $K = \mathrm{rank}(\mathrm{Toep}(\mathbf{v}))$, and where $\bm{\gamma} = [0 \; 1 \; 3 \; 4 \; 9 \; 12 \; 16]^T$.

\subsection{DOA Extraction}
\label{sec:doa}

After solving the fast primal SDP~\eqref{ssdp_fast_mmv} by an off-the-shelf SDP solver (e.g., CVX \cite{grant2014cvx}) and obtaining $\mathbf{u}$, we propose to extract the DOAs by exploiting the IVD of the irregular Toeplitz $\mathrm{T}(\mathbf{u}) = \mathbf{W}(\bm{\gamma}, \mathbf{z}) \mathbf{D} \mathbf{W}(\bm{\gamma}, \mathbf{z})^H$. Although this factorization is not computed explicitly as part of solving the SDP, its existence provides a means to estimate the entries of $\mathbf{z}$, each corresponding to a point on the unit circle whose complex angle encodes a DOA.

Let $\mathbf{T}(\mathbf{u})$ denote an $(N_u,K)$-irregular Toeplitz matrix that has an IVD of the form $\mathrm{T}(\mathbf{u}) = \mathbf{W}(\bm{\gamma}, \mathbf{z}) \mathbf{D} \mathbf{W}(\bm{\gamma}, \mathbf{z})^H$, where $\bm{\gamma} = [\mathcal{U}_1  \dots \mathcal{U}_{N_u}]^T$. Consider the eigen-decomposition of $\mathbf{T}(\mathbf{u})$:
\begin{equation}
    \mathbf{T}(\mathbf{u}) = \mathbf{U}_S \mathbf{\Lambda}_S \mathbf{U}_S^H + \mathbf{U}_N \mathbf{\Lambda}_N \mathbf{U}_N^H,
\end{equation}
where $\mathbf{\Lambda}_S \in \mathbb{C}^{K \times K}$ is a diagonal matrix containing the $K$ largest eigenvalues of $\mathbf{T}(\mathbf{u})$, $\mathbf{U}_S \in \mathbb{C}^{N_u \times K}$ contains the corresponding eigenvectors, and $\mathbf{\Lambda}_N \in \mathbb{C}^{(N_u - K) \times (N_u - K)}$ and $\mathbf{U}_N \in \mathbb{C}^{N_u \times (N_u - K)}$ contain the remaining (zero) eigenvalues and corresponding eigenvectors. $\mathbf{U}_S$ and $\mathbf{U}_N$ are known as the \textit{signal and noise subspaces}, respectively. 

\begin{figure}[!t]
\centering
\includegraphics[width=8.5cm]{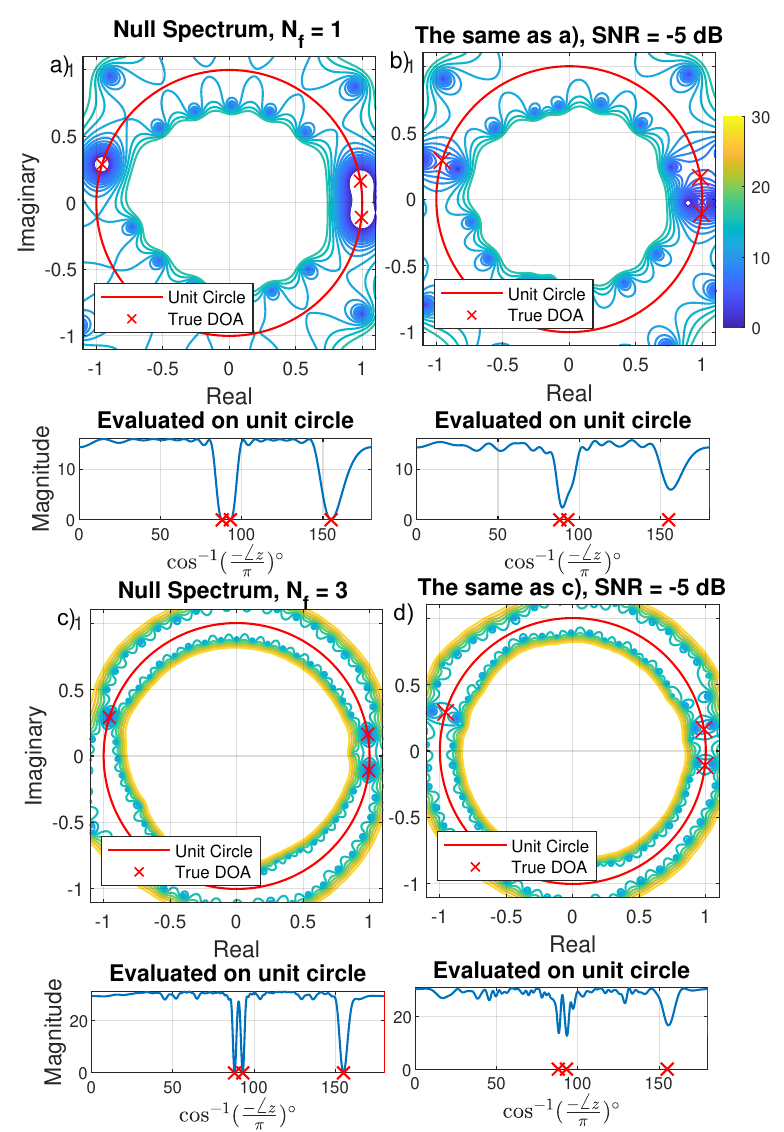}
\caption{Null spectrum contours (dB) using $N_M = 16$, $K = 3$, $N_l = 5$. For a) - b) $N_F = 1$, and for c) - d) $N_F = 3$. The frequency set is $\{100, \dots, N_F \cdot 100\}$ Hz and ULA is applied. DOAs are $[88, 93, 155]^\circ$, marked by red x's, the red line marks the complex unit circle. a) and c): Null spectrum from noise-free measurement. b) and d): Null spectrum from SNR = $-$5 dB measurement.  }
\label{null_spec}
\end{figure}

For $z \in \mathbb{C}$, define the irregular null spectrum $\Tilde{D}(z)$ of $\mathbf{T}(\mathbf{u})$ as \cite[eq. (29)]{wagner2021}
\begin{equation}
\label{D_z}
    \Tilde{D}(z) = \mathbf{w}(\bm{\gamma}, z)^H \mathbf{U}_N \mathbf{U}_N^H \mathbf{w}(\bm{\gamma}, z) = \mathbf{w}(\bm{\gamma}, z)^H \mathbf{G} \mathbf{w}(\bm{\gamma}, z), 
\end{equation}
where $\mathbf{G} = \mathbf{U}_N \mathbf{U}_N^H$. The behavior of the irregular null spectrum is plotted in Fig. \ref{null_spec}. 

Since $\mathbf{G} \perp \mathbf{W}(\bm{\gamma}, \mathbf{z})$ and $|\mathbf{z}| = 1$, the DOAs encoded in $\mathbf{z}$ are associated to the $K$ roots of $\Tilde{D}(z)$ on the unit circle. \cite{wagner2021} suggests that the local minima of $\Tilde{D}(z)$ evaluated on the unit circle give DOA estimates with similar accuracy as those given by the actual roots. Therefore, $\mathbf{z}$ is estimated as \cite[eq. (43)]{wagner2021}
\begin{equation}
   \hat{\mathbf{z}} = \mathrm{arg} \min_{|z| = 1}^k \Tilde{D}(z), \quad k = 1, \dots, K
\end{equation}
where $\mathrm{arg} \min_z^k$ denotes the argument, $z$, which produces the $k$th smallest local minima. The DOAs $\hat{\theta}, \hat{w},$ and $\hat{z}$ are estimated by
\begin{equation}
\hat{\theta} = \cos^{-1}\bigg(-\frac{\angle \hat{z}}{\pi} \bigg), \quad
\hat{w} = \frac{-\angle \hat{z}}{2 \pi}, \quad
\hat{z} = e^{-j \pi \cos(\hat{\theta})}.
\end{equation}

In summary, we first solve the SDP \eqref{ssdp_fast_mmv} via an off-the-shelf SDP solver (e.g., CVX \cite{grant2014cvx}). After $\mathbf{u}$ is obtained, the DOAs can be retrieved by computing the irregular null spectrum $\Tilde{D}(z)$ of $\mathrm{T}(\mathbf{u})$ and following the steps mentioned in this section. The implementation details of the proposed method are summarized in Algorithm \ref{ag}. \\

\algnewcommand\INPUT{\item[\textbf{Input:}]}%
\algnewcommand\OUTPUT{\item[\textbf{Output:}]}%
\algnewcommand\Initialize{\item[\textbf{Initialization:}]}
\begin{algorithm} 
\caption{Regularization-free DOA estimation}\label{ag}
\begin{algorithmic}
\INPUT $\mathcal{Y} \in \mathbb{C}^{N_m \times N_l \times N_f}$, $K$
\Initialize 
\State Solve \eqref{ssdp_fast_mmv} by CVX and obtain $\mathbf{u}$ 
\State Obtain $\mathrm{T}(\mathbf{u})$ based on \eqref{T_map}
\State [$\mathbf{U}$, $\mathbf{\Lambda}$] = $\mathrm{eig}$($\mathrm{T}(\mathbf{u})$)
\State $\mathbf{U}_N = \mathbf{U}(:, K + 1 : N_u)$
\State $\mathbf{G} = \mathbf{U}_N \mathbf{U}_N^H$
\State Obtain $\Tilde{D}(z)$ based on \eqref{D_z}
\State $\hat{\mathbf{z}} = \mathrm{find}(\mathrm{arg} \min(\Tilde{D}(z)), |\mathbf{z}| = 1)$
\State $\hat{\mathbf{\theta}} \gets 180 - \mathrm{acosd}(\mathrm{angle}(\hat{\mathbf{z}} /\pi))$
\OUTPUT $\hat{\mathbf{\theta}}$
\end{algorithmic}
\end{algorithm}

\section{More Sources Than Sensors for ULA}
\label{ula_source}
Many prior works have demonstrated the possibility of resolving more sources than the number of array sensors based on special array geometries such as MRA \cite{moffet1968minimum, zhou2018direction}, co-prime arrays \cite{vaidyanathan2010sparse}, and nested array \cite{pal2010nested}. However, for single-frequency ULA, the maximum number of resolvable sources is $N_M - 1$ \cite[Sec 11.2.3]{yang2018sparse}. In this section, we will demonstrate the possibility of resolving more sources than sensors \textit{under the ULA setup} if multiple frequencies are available. We primarily solve \eqref{ssdp_mmv} and follow the procedures in Algorithm \ref{ag} to retrieve the DOAs. In our multi-frequency ANM configuration, it can resolve up to $N - 1 = (N_M - 1)N_F$ sources as $\mathrm{Toep}(\mathbf{u}) \in \mathbb{C}^{N \times N}$ and $\mathbf{U}_N$ exists only if $K \leq N - 1 $. The reason for using \eqref{ssdp_mmv} instead of \eqref{ssdp_fast_mmv} is that \eqref{ssdp_fast_mmv} can resolve up to $N_u - 1$ sources and \eqref{ssdp_mmv} has the potential to resolve more sources than \eqref{ssdp_fast_mmv} because $\mathrm{Toep}(\mathbf{u})$ in \eqref{ssdp_mmv} has a higher dimension than $\mathrm{T}(\mathbf{u})$ in \eqref{ssdp_fast_mmv}. \textcolor{black}{This idea was also demonstrated in \cite{qin2017doa} for co-prime frequencies, though the method in \cite{qin2017doa} used grid-based DOA estimation.}

The key observation for the multi-frequency model is that these frequencies increase the diversity of the harmonics. These extra harmonics serve as ``virtual'' sensors in the array, and they bring about an enhanced degree of freedom. For example, consider a ULA with $N_M = 4$ sensors and $N_F = 5$ uniform frequencies. Therefore, it can resolve up to $(N_M - 1)N_F = 15$ sources. The SDP problem \eqref{ssdp_fast_mmv} can be interpreted as a structured covariance matrix estimation problem ($\mathrm{T}(\mathbf{u})$ can be interpreted as the covariance matrix). We notice this covariance matrix is in a higher dimension, which corresponds to our intuition that there are more sensors in our ``virtual'' array. 

\begin{figure}[!t]
\centering
\includegraphics[width=8.5cm]{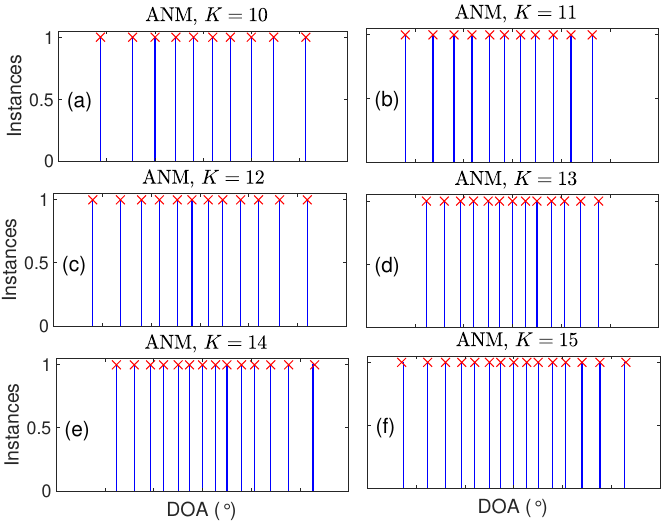}
\caption{Estimated and True DOAs for ANM (``$\times$'' indicates the true DOAs and the black vertical line indicates the estimated DOAs). $N_M = 4$, $N_F = 5$, $N_l = 1$, and $K = 10, 11, 12, 13, 14, 15$. The RMSEs of ANM under $K = 10, 11, 12, 13, 14, 15$ are $0.005^\circ$, $0.16^\circ$, $0.20^\circ$, $0.04^\circ$, $0.27^\circ$, and $0.27^\circ$.}
\label{hist_nf}
\end{figure}

As an example, suppose we have $N_M = 4$ sensors, $N_F = 5$ frequencies ($\{$100, \dots, 500$\}$ Hz), $N_l = 1$ noise-free snapshot, and $K = 10, 11, 12, 13, 14, 15$ sources with uniform and deterministic across  frequencies. For $K = 10, 12, $ and $15$, the DOAs are generated as the uniform distribution in the cosine domain (i.e., the DOAs are $\lfloor \cos^{-1}(-1 + 2([1:K] - 0.5)/K) \rfloor$). For $K = 11$, we pick up the last $11$ sources in the $K = 12$ case. For $K = 13$, we pick up the middle $13$ sources in the $K = 15$ case, and for $K = 14$, we pick up the middle $14$ sources.  We plot the estimated DOAs for ANM. From Fig. \ref{hist_nf}, we can see our ANM can resolve up to $(N_M - 1)N_F = 15$ sources.

\section{Rank Minimization and Atomic $\ell_0$ Norm Minimization} 
\label{sec:rankdiscussion}
In this section, we highlight the connection between rank minimization and atomic $\ell_0$ norm minimization. More specifically, atomic $\ell_0$ norm minimization can be interpreted as a covariance matrix estimation approach where low-rankness and Toeplitz structure are explicitly enforced \cite{tang2013compressed, yang2016exact, liu2021rank}. However, atomic $\ell_0$ norm minimization is non-convex and may not be computationally feasible. By considering ANM, the convex relaxation of the atomic $\ell_0$ norm, we obtain~\eqref{ssdp_fast_mmv} and~\eqref{ssdp_fast_smv} as trace minimization problems that are computationally feasible and in which the low-rankness and Toeplitz structure are implicitly enforced. In this way, we can understand the benefits of ANM compared to conventional covariance matrix estimation using the sample covariance matrix. Before describing the equivalence, we review the definitions of the covariance matrix and the sample covariance matrix. 

We assume $N_l = 1$, noise-free measurement, uniform frequency and ULA setup, and full dimensional SDP in this section, and our discussion serves as a means to interpret~\eqref{ssdp_fast_smv}. 

\subsection{Covariance Matrix Estimation}
\label{sec:covmatest}

Suppose $\mathbf{\widetilde{Y}} \in \mathbb{C}^{N \times N_F}$ is noise-free and defined as
\begin{equation}
    \mathbf{\widetilde{Y}} = \sum_{k = 1}^K \mathbf{z}_k \mathbf{x}_k^H  = [\mathbf{z}_1 \dots \mathbf{z}_K] [\mathbf{x}_1 \dots \mathbf{x}_K]^H = \mathbf{ZX}
    \label{eq:ytildedef}
\end{equation}
where $\mathbf{z}_k := [z_k^{0} \dots z_k^{N-1}]^T \in \mathbb{C}^{N}$, 
$\mathbf{x}_k = [x_k^{(1)} \dots x_k^{(N_F)}]^T \in \mathbb{C}^{N_F}$, 
$\mathbf{Z} := [\mathbf{z}_1 \dots \mathbf{z}_K] \in \mathbb{C}^{N \times K}$, and $\mathbf{X} := [\mathbf{x}_1 \dots \mathbf{x}_K]^H  \in \mathbb{C}^{K \times N_F}$. Note that $\mathbf{Z}$ is a Vandermonde matrix and that $\mathbf{\widetilde{Y}}$ satisfies
\begin{equation}
    \mathbf{Y} = \mathcal{R}^*(\widetilde{\mathbf{Y}}).
\end{equation}

The sample covariance matrix $\hat{\mathbf{R}}_{\tilde{y}\tilde{y}}$ and covariance matrix $\mathbf{R}_{\tilde{y}\tilde{y}}$ are defined as 
\begin{equation}
\label{sample_cov}
    \hat{\mathbf{R}}_{\tilde{y}\tilde{y}} = \frac{1}{N_f} \mathbf{\widetilde{Y}} \mathbf{\widetilde{Y}}^H,
\end{equation}
\begin{equation}
\label{cov_est}
    \mathbf{R}_{\tilde{y}\tilde{y}} = \EX[\hat{\mathbf{R}}_{\tilde{y}\tilde{y}}] = \mathbf{Z} \mathbf{\Lambda}_X \mathbf{Z}^H 
\end{equation}
where $\mathbf{\Lambda}_X := \frac{1}{N_f} \EX[\mathbf{XX}^H]$ is a diagonal matrix due to uncorrelated sources. Note that $\mathbf{R}_{\tilde{y}\tilde{y}}$ is a Toeplitz Hermitian matrix in the noise-free case, $\mathbf{Z} \mathbf{\Lambda}_X \mathbf{Z}^H $ is its Vandermonde decomposition, and the DOAs are encoded in the Vandermonde matrix $\mathbf{Z}$. The essence of some classical DOA estimation approaches (e.g. MUSIC, and ESPRIT) lies in Vandermonde decomposition of the estimated covariance matrix. 

However, $\mathbf{\widetilde{Y}}$ is not fully observed in our problem as only $\mathbf{Y}$, the image of the $\mathcal{R}^*$ mapping is accessible. To further obtain $\tilde{\mathbf{Y}}$, the \textcolor{black}{lifting mapping $\mathcal{R}$} needs to be applied. Note if we apply $\mathcal{R}^*$ first and then $\mathcal{R}$ on a matrix, we may not obtain the same matrix as the white entries in Fig. \ref{R_map_plot} cannot be recovered after the $\mathcal{R}^*$ mapping. Therefore, the covariance matrix of $\tilde{\mathbf{Y}}$ must be estimated by solving a convex optimization problem.

\subsection{Connection Between Rank Minimization and Atomic $\ell_0$ Norm Minimization}

As described in Sec.~\ref{sec:doa}, after solving the SDP \eqref{ssdp_fast_smv} and obtaining $\mathbf{u}$, DOAs are extracted by computing the IVD of $\mathrm{T}(\mathbf{u})$. In light of the discussion in Section~\ref{sec:covmatest}, then, \eqref{ssdp_fast_smv} can be interpreted as a covariance matrix estimation problem where $\mathrm{T}(\mathbf{u})$ serves as an estimate for a covariance matrix that contains the DOA information. In this section, we discuss this connection more deeply.

From \eqref{cov_est}, the true covariance matrix in the noise-free case $ \mathbf{R}_{\tilde{y}\tilde{y}}$ has three important properties: (1) Toeplitz and Hermitian; (2) PSD; (3) low-rank (its rank is $K$ (number of sources) and is usually much smaller than its size $N_u$). A commonly used estimate for the covariance matrix is the \textit{sample covariance matrix} $\hat{\mathbf{R}}_{\tilde{y}\tilde{y}}$ defined in \eqref{sample_cov}, which is PSD. However, this estimate does not promote the Toeplitz structure of the covariance matrix. This limitation is overcome by the SDP formulation in \eqref{ssdp_fast_smv}. The irregular Toeplitz structure is obviously enforced in $\mathrm{T}(\mathbf{u})$. Meanwhile, \eqref{ssdp_fast_smv} also promotes low-rank structure, a fact that warrants more discussion.

The atomic $\ell_0$ norm of an $N \times N_F$ matrix $\mathbf{\widetilde{Y}}$ is defined as 
\begin{equation}
    \label{eq:an0def}
    \|\mathbf{\widetilde{Y}} \|_{\mathcal{A}, 0} := \inf \Bigg\{K \Bigg| \mathbf{\widetilde{Y}} = \sum_{k = 1}^K c_k\mathbf{z}_k \mathbf{x}_k^H, c_k > 0  \Bigg\}
\end{equation}
where $\mathbf{z}_k  := [z_k^{0} \dots z_k^{N-1}]^T \in \mathbb{C}^{N}$ such that $|z_k| = 1$ and 
$\mathbf{x}_k = [x_k^{(1)} \dots x_k^{(N_F)}]^T \in \mathbb{C}^{N_F}$ such that $\|\mathbf{x}_k\|_2 = 1$.

The following proposition establishes an equivalence between the atomic $\ell_0$ norm and rank minimization. 
\begin{proposition}(\cite[Theorem 11.13]{yang2018sparse})
For any $N \times N_F$ matrix $\mathbf{\widetilde{Y}}$ with an atomic decomposition of the form \eqref{eq:an0def} (which includes any $\mathbf{\widetilde{Y}}$ satisfying~\eqref{eq:ytildedef}), $\|\mathbf{\widetilde{Y}} \|_{\mathcal{A}, 0}$ is equal to the optimal value of the following rank minimization problem:
\begin{equation}
\label{rank_min}
\begin{aligned}
&\min_{\mathbf{W}, \mathbf{u}} \quad \mathrm{rank}(\mathrm{Toep}(\mathbf{u})) \\
    \quad &\textrm{s.t.} 
\left[                 
  \begin{array}{cc}   
    \mathrm{Toep}(\mathbf{u}) & \mathbf{\widetilde{Y}} \\  
    \mathbf{\widetilde{Y}}^H & \mathbf{W} \\  
  \end{array}
\right]  \succeq 0.
\end{aligned}
\end{equation}
\end{proposition}

\textbf{Remark} To summarize our intuition, the proposition above indicates that \eqref{ssdp_fast_smv}, which is essentially a reduced-dimension convex relaxation of \eqref{rank_min}, will promote both low-rankness and Toeplitz structure and therefore yields a favorable covariance matrix estimation that reveals the sparse decomposition of the DOAs and is consistent with the observed data.

\section{Numerical Results}
We use numerical experiments to examine the performance of the method. In this section, $N_F$ and $N_f$ are used to denote the number of frequencies for the uniform and non-uniform frequency set, respectively. $N_M$ and $N_m$ are used to denote the number of sensors for the uniform and non-uniform array spacing sets, respectively. For each experiment and trial, $K$ DOAs are generated. The source amplitude is complex Gaussian. $N_l$ snapshots are collected. The uniform frequency set is defined as $\{1, ..., N_F\} \cdot F_1$ ($F_1$ is the minimum frequency). The array spacing for ULA is $\frac{\lambda_1}{2}$ where $\lambda_1$ is the wavelength for the minimum frequency in the frequency set.  The noise for each frequency and each snapshot is randomly generated from the complex Gaussian distribution $\mathcal{CN}(0, \sigma^2)$ and then scaled to fit the desired signal-to-noise ratio (SNR) defined as 
\begin{equation}
    \mathrm{SNR} = 20 \log_{10} \frac{\| \mathcal{X} \|_{\mathrm{HS}}}{\| \mathcal{N} \|_{\mathrm{HS}}}.
\end{equation}

In the Monte-Carlo experiments, $MC = 100$ trials are executed to compute the root mean square error (RMSE) defined as 
\begin{equation}
\label{rmse}
    \mathrm{RMSE} = \sqrt{\frac{1}{MC} \sum_{m = 1}^{MC}  \bigg[ \min \bigg( \frac{1}{K}\sum_{k = 1}^{K}(\hat{\theta}_{mk} - \theta_{mk})^2, 10^2 \bigg) \bigg]},
\end{equation}
where $\hat{\theta}_{mk}$, and $\theta_{mk}$ are (sorted) estimated DOAs, and (sorted) ground-truth DOAs for the $k$th DOA and $m$th trial. A maximum threshold of $10^\circ$ is used to penalize the incorrect DOA estimates. We compare the proposed method with the multi-frequency sparse Bayesian learning (SBL) \cite{nannuru2019sparse}. The Cram{\'e}r-Rao bound (CRB) \cite[Eq. (121)]{liang2020review} for the multi-frequency model is computed for reference. 

\subsection{Robustness to Aliasing/Collision}
\label{aliasing_robust}
We first examine the robustness of aliasing/collision. Suppose $K = 3$ sources impinge in a ULA with $N_M = 16$ sensors. The source amplitudes are complex Gaussian and the DOAs are randomly generated from a uniform distribution with range $[15^\circ, 165^\circ]$ with minimum separation $4/N_M$ in the cosine  \textcolor{black}{ domain inspired by \cite[Theorem 4.2]{wu2023gridless}}. We consider $N_F = 2$ or $4$ under the single-snapshot and uniform frequency case ($N_l = 1$). All frequencies other than the fundamental frequency will have the risk of aliasing/collision. We solve the SDP program \eqref{ssdp_fast_smv} by CVX \cite{grant2014cvx} and apply the root-MUSIC (Vandermonde decomposition) to retrieve the DOAs.

From Fig.\ \ref{RMSE_4plot} (a)--(b), the primal ANM (ANM P) is more robust to the aliasing than SBL. It also overcomes the collision issues for the dual ANM (ANM D)\cite{wu2023gridless}. Moreover, the primal ANM does not need any hyper-parameter tuning and it avoids the bias from the regularization terms.

\textcolor{black}{
\subsection{Non-uniform Power}
In the previous section, the power of each source was the same (all $c_w=1$). In this section, we examine the case when the power of each source is different. From Fig.\ \ref{RMSE_power}, the proposed method can achieve almost the same performance as the uniform power case and therefore it can be applied to the case when the power for each source is different.
}

\subsection{Non-uniform Frequency Cases}
We examine the performance under the non-uniform frequency set. In this case, $N_f = 4$, and the frequency set is $\{100, 200, 300, 500 \}$ Hz and $\{200, 300, 400, 500\}$ Hz. Other conditions are the same as in Sec. \ref{aliasing_robust}. Fig. \ref{RMSE_4plot}~(c)-(d) demonstrates the effectiveness of the proposed method under the non-uniform frequency case. We see superior performance to the fast dual algorithm proposed in~\cite{wu2023gridless}.

\subsection{MMV Case}
\begin{figure}[!t]
\centering
\includegraphics[width=8.5cm]{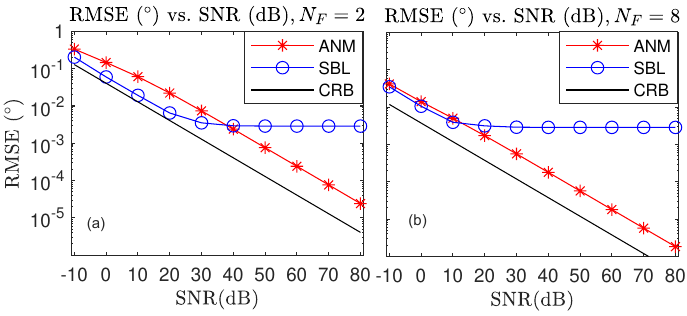}
\caption{RMSE ($^\circ$) versus SNR for MMV setup. $N_M = 16$ ULA with $d = \lambda_{100}/2$. $K = 3$ DOAs at $[88^\circ, 93^\circ, 155^\circ] + \bm{\epsilon}$ where $\bm{\epsilon}$ is the random offsets from a uniform distribution $[0, 1]$. $N_l = 20$. (a): $N_F = 2$ with frequency set $\{100, 200\}$ Hz; (b) $N_F = 8$ with frequency set $\{100, \dots, 800\}$ Hz. }
\label{RMSE_mmv}
\end{figure}

We examine the performance of ANM under the MMV setup. We consider the case $N_l = 20$, and $K = 3 $ DOAs at $[88, 93, 155]^\circ + \bm{\epsilon}$ where $\bm{\epsilon}$ is a three dimensional random vector with uniform distribution from $[0, 1]$. Fig. \ref{RMSE_mmv} demonstrates the superior performance of ANM in the high SNR region, and it follows the trend of CRB. 

\begin{figure}[!t]
\centering
\includegraphics[width=8.5cm]{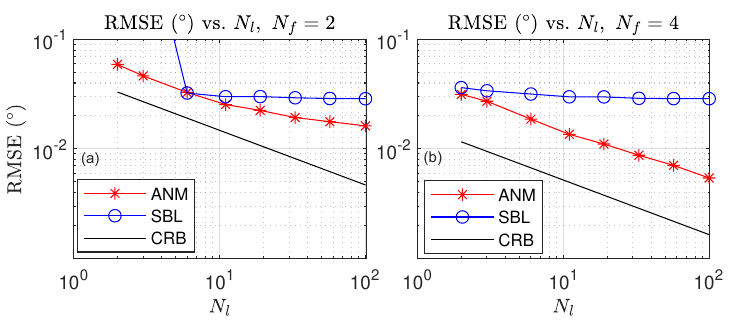}
\caption{RMSE ($^\circ$) versus $N_l$ for MMV setup. $N_M = 16$ ULA with $d = \lambda_{100}/2$. $K = 3$ DOAs at $[88^\circ, 93^\circ, 155^\circ] + \bm{\epsilon}$ where $\bm{\epsilon}$ is the random offsets from a uniform distribution $[0, 1]$. SNR = 20 dB. (a): $N_F = 2$ with frequency set $\{100, 200\}$ Hz; (b) $N_F = 4$ with frequency set $\{100, 200, 300, 400 \}$ Hz. }
\label{RMSE_nl}
\end{figure}

We then examine the performance of ANM with varying numbers of snapshots $N_l$ for SNR = 20 dB, and the other setup as Fig. \ref{RMSE_mmv}.  From Fig. \ref{RMSE_nl}, we can see ANM follows the trend of CRB and outperforms SBL. In addition, comparing Fig. \ref{RMSE_nl} (a) with Fig. \ref{RMSE_nl} (b), ANM performs better with higher $N_F$, which demonstrates the benefits of muli-frequency processing. 

\subsection{The Effect of Multiple Frequencies}
\begin{figure}[!t]
\centering
\includegraphics[width=8.75cm]{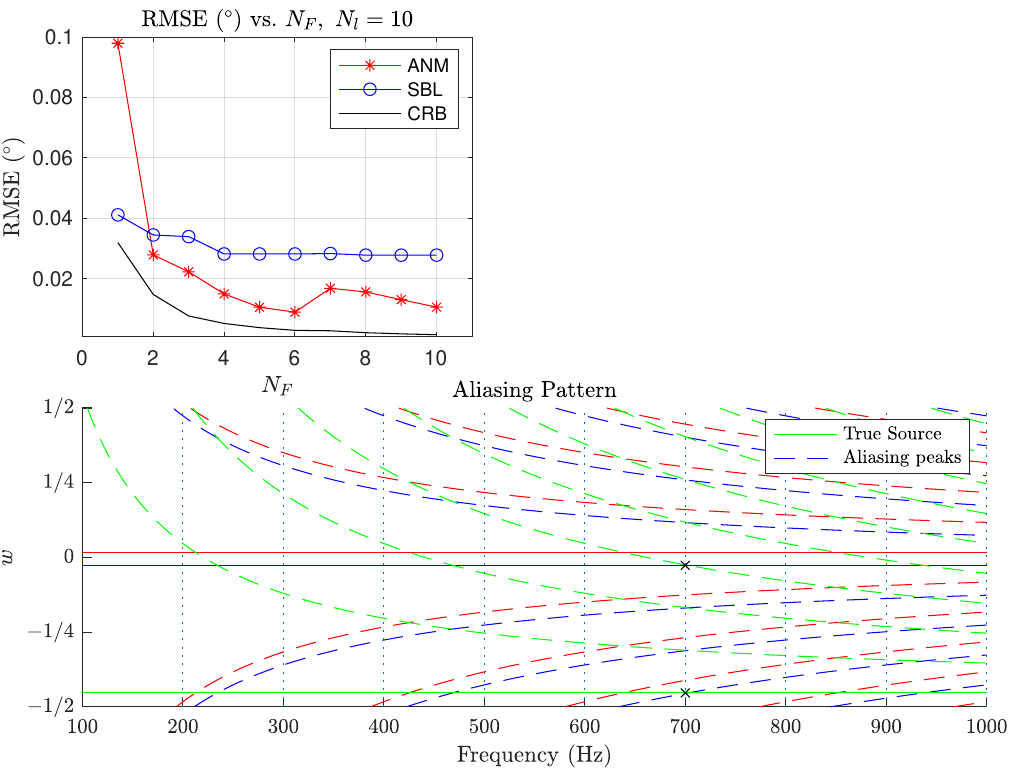}
\caption{(a) RMSE ($^\circ$) versus $N_F$ and (b) aliasing pattern for MMV setup. $N_M = 16$ ULA with $d = \lambda_{100}/2$. $N_l = 10$. The frequency set is $\{100, \dots, N_F \cdot 100\}$ Hz. $K = 3$ DOAs at $[88^\circ, 93^\circ, 155^\circ] + \bm{\epsilon}$ where $\bm{\epsilon}$ is the random offsets from a uniform distribution $[0, 1]$. SNR = 20 dB.  In (b), the true (solid) and the aliasing DOAs (dashed) are shown, with true DOAs [$88^\circ$ (red), $93^\circ$ (black), $155^\circ$ (green)].}
\label{RMSE_nf}
\end{figure}

We study the performance of the method under varying $N_F$ in Fig. \ref{RMSE_nf}. From Fig. \ref{RMSE_nf} (a), the estimation error of ANM generally goes down with increasing $N_F$ and the only exception is $N_F = 7$, \textcolor{black}{where it increases by roughly $0.01^\circ$}. To understand that, the true and the aliasing DOAs are in Fig.\ \ref{RMSE_nf} (b). It can be seen that the DOAs $93^\circ$ and $155^\circ$ \textcolor{black}{nearly} collide with each other at frequency $700$ Hz. \textcolor{black}{Referring to \eqref{eq:w} and \eqref{eq:nearcollisson}, this can be understood as $w_2 = 1/2 \cdot \cos(93^\circ)$, $w_3 = 1/2 \cdot \cos(155^\circ)$, $|w_2 - w_3| \approx 3/7$, and there is a near collision in frequency $f=7$ (i.e. $7 \cdot 100 = 700$ Hz).} There are other intersection points between the solid and dashed lines but none of them lie in any frequency that belongs to the frequency set. That explains why the error increases when $N_F$ increases from 6 to 7.

\begin{figure}[!t]
\centering
\includegraphics[width=8.5cm]{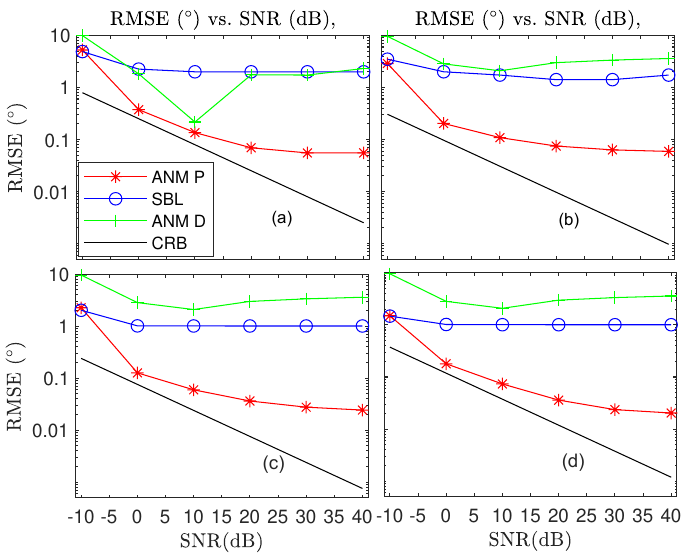}
\caption{RMSE ($^\circ$) versus SNR. $N_M = 16$ ULA with $d = \lambda_{100}/2$. $N_l = 1$, and $K = 3$ sources with randomly generated DOAs from $[15^\circ, 165^\circ]$ with minimum separation $4/N_M$ in the cosine domain. (a): $N_F = 2$ with frequency set $\{100, 200\}$ Hz; (b) $N_F = 4$ with frequency set $\{100, 200, 300, 400 \}$ Hz; (c) $N_f = 4$ with frequency set $\{100, 200, 300, 500 \}$ Hz; (d) $N_f = 4$ with frequency set $\{200, 300, 400, 500 \}$ Hz. The proposed primal SDP program (ANM P) and the dual SDP program \cite{wu2023gridless} (ANM D) as well as SBL and CRB are shown. All source amplitudes are complex Gaussian with unit variance and $c_w = 1$.}
\label{RMSE_4plot}
\end{figure}

\begin{figure}[!t]
\centering
\includegraphics[width=8.5cm]{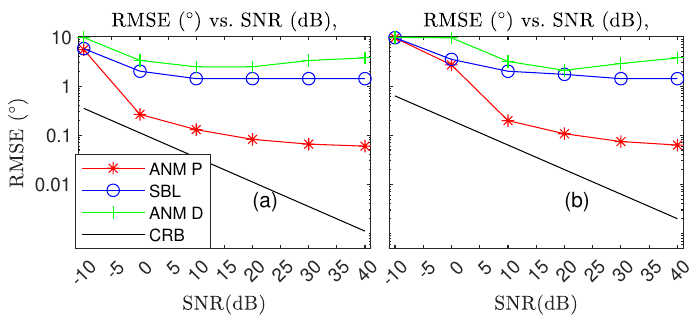}
\caption{\textcolor{black}{Same setup as Fig. \ref{RMSE_4plot} (b), but the powers of the three sources are (a)~1, 2, 3  and (b)~1, 4, 16.} }
\label{RMSE_power}
\end{figure}

\begin{figure}[!t]
\centering
\includegraphics[width=8.5cm]{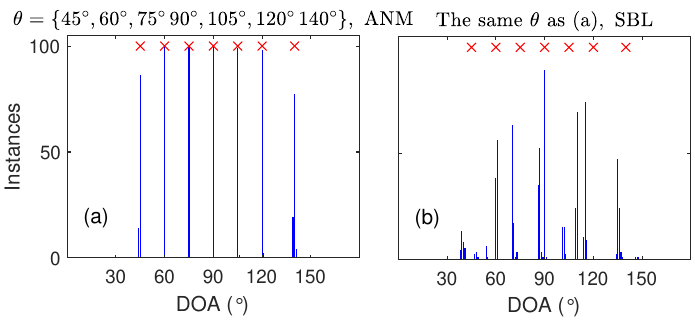}
\caption{Histogram for the estimated DOAs for (a) ANM, and (b) SBL under the co-prime array with $N_m $= 6 (sensor locations are $[0, 2, 3, 4, 6, 9]$), $N_f$ = 3 ($[100, 300, 400]$ Hz), $N_l$ = 50, SNR = $20$ dB, and $K$ = 7. The RMSE for ANM is $0.2^\circ$, and for SBL $8.6^\circ$.}
\label{RMSE_nua}
\end{figure}

\begin{figure}[!t]
\centering
\includegraphics[width=8.5cm]{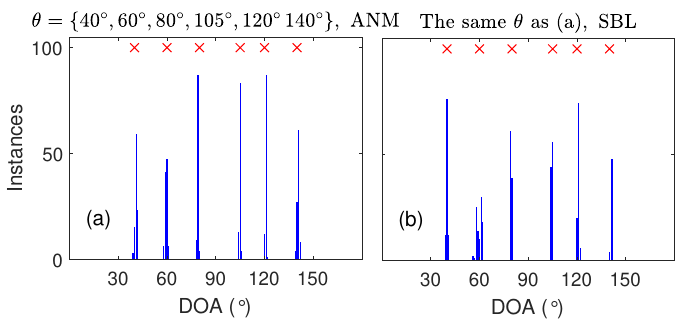}
\caption{Histogram for the estimated DOAs for an ULA (a) ANM, and (b) SBL. $N_M = 4$, $N_F = 3$ ($[100, 200, 300]$ Hz), $N_l = 50$, SNR = 20 dB and $K = 6$. The RMSE for ANM is $0.90^\circ$, and for SBL  $1.10^\circ$.}
\label{hist_coprime}
\end{figure}

\begin{figure}[!t]
\centering
\includegraphics[width=4.5cm]{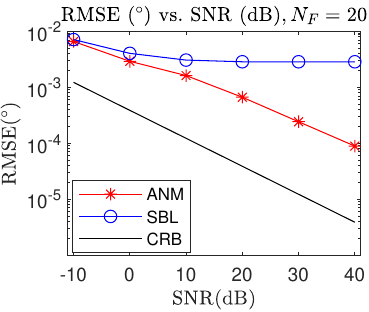}
\caption{RMSE ($^\circ$) versus SNR for MMV setup. $N_M = 20$ ULA with $d = \lambda_{100}/2$. $K = 3$ DOAs at $[88^\circ, 93^\circ, 155^\circ] + \bm{\epsilon}$ where $\bm{\epsilon}$ is the random offsets from a uniform distribution $[0, 1]$. $N_l = 10$. (a): $N_F = 20$ with frequency set $\{100, \dots, 2000\}$ Hz. }
\label{Nf_20}
\end{figure}

\subsection{Co-prime Array and More Sources than Sensors}
We examine an $N_m  = 6$ co-prime array, a particular example of the non-uniform array (NUA). A co-prime array involves two ULAs with spacing $M_1d$ and $M_2d$. $M_1$ and $M_2$ are co-prime integers and their greatest common divisor is 1. The first ULA has $M_2$ sensors and the second ULA has $2M_1$ sensors. Since the first sensor is shared, there are $N_m = 2M_1 + M_2 - 1$ sensors in the array.  In this example, we consider $M_1 = 2$, $M_2 = 3$. $N_f = 3$ and the non-uniform frequency set is $\{100, 300, 400\}$ Hz. $d = \lambda_{100} / 2$. The first ULA is $[0, 2d, 4d]$ and the second ULA is $[0, 3d, 6d, 9d]$. The entire co-prime array is $[0, 2d, 3d, 4d, 6d, 9d]$. $N_l = 50$, SNR = $20$ dB, and $K = 7$ DOAs with at $\{45, 60, 75, 90, 105,  120, 140 \}^\circ$. Note, $K > N_m$ in this case. 

From Fig. \ref{RMSE_nua}, the proposed method resolves more DOAs than sensors in the NUA case, while SBL fails in this case and has a high RMSE (The maximum RMSE is $10^\circ$ as the maximum threshold of the RMSE for one trial is $10^\circ$ based on \eqref{rmse}).

Further, we examine the case when there are more DOAs than sensors \textit{under the ULA setup}. 
We have already demonstrated that in Sec. \ref{ula_source} under a noise-free and uniform amplitude setup. Here, we consider a more practical case when there is noise and the amplitude is random. From Fig. \ref{hist_coprime}, ANM can resolve $6$ DOAs when only $N_M = 4$ physical sensors are available under the noisy and non-uniform amplitude case and it achieves lower RMSE performance than SBL. 

\textcolor{black}{Although we only demonstrated the co-prime array as an important example of NUA, the  proposed method can be applied to any NUA satisfying the assumptions in Sec. \ref{assump}-1.}

\subsection{Practical Test}
We consider a case with $N_F = 20$ frequencies, $N_l = 10$ snapshots, and $N_M = 20$ sensors. In previous examples, the number of frequencies is small, but in practical cases, there may be many more frequencies in the wideband signal. We compare the performance to the SBL with high resolution $0.01^\circ$. From Fig. \ref{Nf_20}, our method can deal with such a practical case with lower RMSE than the high-resolution SBL. 

\begin{figure}[!t]
\centering
\includegraphics[width=8.5cm]{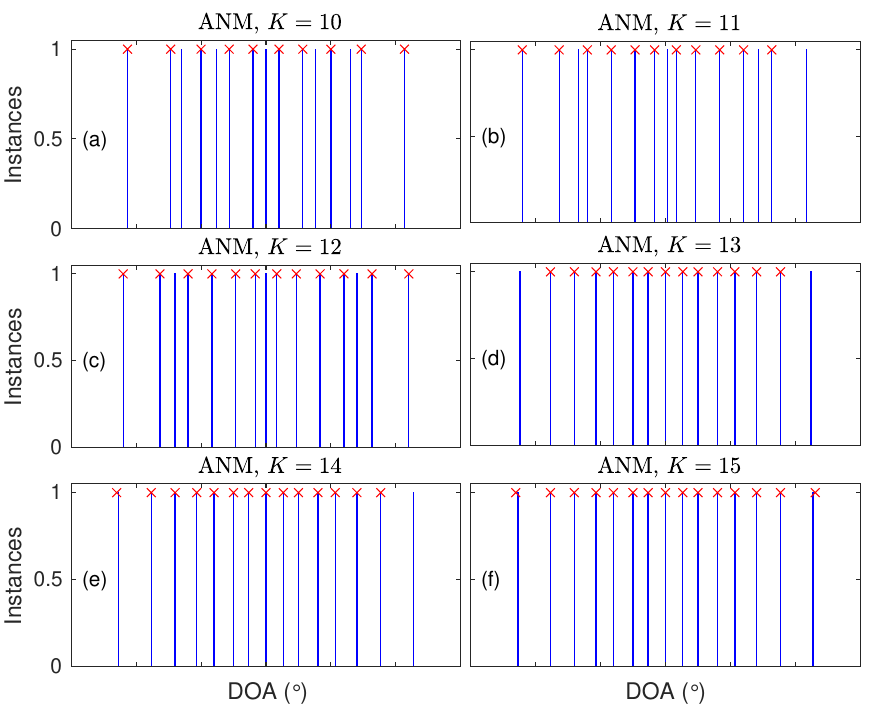}
\caption{Estimated and True DOAs for ANM (``$\times$'' indicates the true DOAs and the black vertical line indicates the estimated DOAs). The same setup as Fig. \ref{hist_nf} except $K$ is estimated as $15$ for all cases. }
\label{hist_K}
\end{figure}

\textcolor{black}{
\subsection{Robustness to Inaccurate Source Number Estimation}
In the previous examples, we all assume a perfect knowledge of the source number $K$. While in the real applications, the accurate prior knowledge of the source number $K$ may not be available. In the following example, we demonstrate the case when $K$ is overestimated (i.e. $K_{\text{est}} \geq K$). }

\textcolor{black}{
The setup is the same as Fig. \ref{hist_nf} except $K_{\text{est}} = 15$. From Fig. \ref{hist_K}, the proposed method can still capture the $K$ true sources for all cases in this example. Therefore, it has robustness to the case when $K$ is overestimated. In practice, if $K$ is unknown, we can feed a reasonably large $K$ to the proposed method and it is still possible to resolve all the true sources. 
}

\textcolor{black}{
\subsection{Complexity}
We compare the average CPU time of the proposed method with the dual ANM \cite{wu2023gridless} and SBL (with a 0.01$^\circ$ grid) over 100 trials with the setup used in Fig. \ref{RMSE_4plot} (b) with SNR values of $-$10, 10, and 30 dB. In these results, presented in Table \ref{tab:table2}, the proposed method is at least 3 times faster than both the dual ANM and SBL. }

\begin{table}[]
\begin{center}
\caption{\label{tab:table2} Average CPU Run Time (s) Under the Same Setup as Fig. \ref{RMSE_4plot} (b) }
\begin{tabular}{ c | c | c | c}
\hline
  & $-$10 dB & 10 dB  &  30 dB \\ 
\hline
Proposed  & 1.1 & 1.1  & 1.0 \\  
ANM \cite{wu2023gridless} & 3.8 & 3.8  & 3.7 \\
SBL   & 7.0 & 4.9  & 4.8  \\
\hline
\end{tabular}
\end{center}
\end{table}

\section{Conclusion}
This paper proposes a gridless DOA estimation method based on \textit{regularization-free} SDP and Vandermonde decomposition. We further extend this framework to MMV, NUA, and non-uniform frequency cases. Under the NUA and non-uniform frequency case, the Toeplitz structure will not hold. However, we demonstrate the possibility of using IVD in these cases, and the existence of IVD is theoretically guaranteed. With the help of multiple frequencies, the method can resolve more sources than the number of physical sensors under the ULA setup. Therefore, multi-frequency processing can reduce \textcolor{black}{the number of physical sensors} and increase the maximum resolvable sources. Numerical results demonstrate the proposed framework is robust to noise and aliasing and can achieve a superior performance under the MMV, NUA, and NUF setup.

\appendix

\subsection{Proof for Proposition \ref{p1}}
\label{p_31}
\textit{Proof} The primal atomic norm $\|\mathcal{X}\|_{\mathcal{A}}$ is expressed in terms of the dual atomic norm $\|\mathcal{Q}\|_{\mathcal{A}}^*$ as
\begin{equation}
\label{x_anm}
    \|\mathcal{X}\|_{\mathcal{A}} = \sup_{\|\mathcal{Q}\|_{\mathcal{A}}^* \leq 1} \langle \mathcal{Q}, \mathcal{X} \rangle_{\mathbb{R}} = \sup_{\|\mathcal{Q}\|_{\mathcal{A}}^* \leq 1} \langle \mathcal{Q}, \mathcal{Y} \rangle_{\mathbb{R}},
\end{equation}
where the last equality is only for the noise-free case. 
For any dual variable $\mathcal{Q}$, we can define the \textit{dual polynomial matrix} $\bm{\Psi}(\mathcal{Q}, w) \in \mathbb{C}^{N_F \times N_l}$ as
\begin{equation}
\label{x_poly}
    \bm{\Psi}(\mathcal{Q}, w) := [\mathbf{Q}_1^H\mathbf{a}(1, w) \dots \mathbf{Q}_{N_F}^H\mathbf{a}(N_F, w)]^T.
\end{equation}

Since each frequency has different array manifold vectors, it is difficult to express $\bm{\Psi}(\mathcal{Q}, w)$ as a matrix multiplication of $\mathcal{Q}$ and a vector. To construct a homogeneous representation for $\bm{\Psi}(\mathcal{Q}, w)$, we will leverage $\mathbf{z}:= [z^0 \dots z^{N-1}]^T \in \mathbb{C}^N$ ($z = z(w):= e^{-j2 \pi w}$), an ensemble of the array manifold, and the matrix $\widetilde{\mathbf{Q}}_f \in \mathbb{C}^{N \times N_l}$ defined as follows \cite[eq. (14)]{wu2023gridless}
\begin{equation}
\begin{aligned}
    \widetilde{\mathbf{Q}}_f(i,l) = \left\{
\begin{array}{ll}
    \mathbf{Q}_f(m, l) \quad   &\mbox{for} \;\; (i, l)=(f \cdot (m-1) + 1, l)\\
    0  & \textrm{otherwise}, 
\end{array}
\right.
\end{aligned}
\end{equation}
or $\widetilde{\mathbf{Q}}_f = \mathcal{R}(\mathbf{Q}_f)$. 
With the help of $\widetilde{\mathbf{Q}}_f$ and $\mathbf{z}$, $\bm{\Psi}(\mathcal{Q}, w)$ has the representation 
\begin{equation}
    \bm{\Psi}(\mathcal{Q}, w) =  [\widetilde{\mathbf{Q}}_1^H\mathbf{z} \dots \widetilde{\mathbf{Q}}_{N_F}^H\mathbf{z}]^T.
\end{equation}

Now, we consider $\|\mathcal{Q}\|_{\mathcal{A}}^*$, which appears in the constraint in \eqref{x_anm}. Recalling that $\|\mathbf{X}_w\|_F = 1$, we have a similar derivation to \cite[eq. (17)]{wu2023gridless}:
\begin{equation}
\label{anmdual}
\begin{aligned}
\|\mathcal{Q}\|^{*}_{\mathcal{A}} &:= \sup_{\|\mathcal{X}\|_{\mathcal{A}} \leq 1} \langle \mathcal{Q, X} \rangle_{\mathbb{R}} = \sup_{\|\mathcal{X}\|_{\mathcal{A}} \leq 1} \langle \mathcal{Q}, \mathbf{A}(w) * \mathbf{X}_w^T \rangle_{\mathbb{R}}\\
&= \sup_{\substack{\mathbf{X}_w \\ w}}  \mathrm{Tr}[\sum_{f = 1}^{N_F} \mathbf{Q}_f^H \mathbf{a}(f, w) \mathbf{x}_w^T(f)]     \\
&= \sup_{\substack{\mathbf{X}_w \\ w}} \mathrm{Tr}[\bm{\Psi}^H \mathbf{X}_w] = \sup_{w} \|\bm{\Psi}(\mathcal{Q}, w)\|_F.
\end{aligned}
\end{equation}

Using \eqref{anmdual}, the condition $\|\mathcal{Q}\|^{*}_{\mathcal{A}} \leq 1$ can be equivalently formulated as an SDP constraint. Construct a similar polynomial as in \cite[eq. (23)]{wu2023gridless}:
\begin{equation}
\begin{aligned}
    R(w) &:= 1 - \| \bm{\Psi}(\mathcal{Q}, w) \|_F^2 = 1 - \mathrm{Tr}[\bm{\Psi}^H(\mathcal{Q}, w)\bm{\Psi}(\mathcal{Q}, w)] \\
    &= 1 - \mathrm{Tr}(\sum_{f = 1}^{N_F}\widetilde{\mathbf{Q}}_f^H \mathbf{zz}^H \widetilde{\mathbf{Q}}_f) = 1 - \sum_{f = 1}^{N_F} \mathbf{z}^H \widetilde{\mathbf{Q}}_f \widetilde{\mathbf{Q}}_f^H \mathbf{z}.
\end{aligned}
\end{equation}
Therefore, $\|\mathcal{Q}\|^{*}_{\mathcal{A}} \leq 1$ holds if and only if $R(w) \geq 0$ for all $w \in [-1/2, 1/2]$.

\begin{figure}[!t]
\centering
\includegraphics[width=8.5cm]{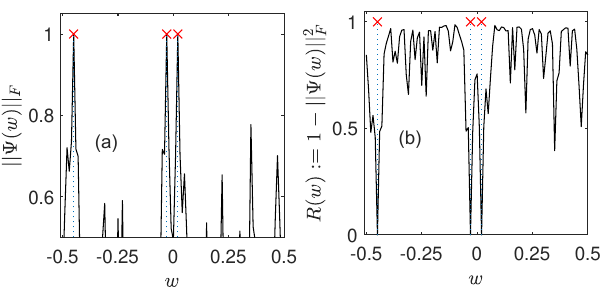}
\caption{Dual polynomial visualization. A ULA with $N_M = 16$ sensors and spacing $d = c/2F_1$ ($F_1 = 100$ Hz) is used. $N_F = 5$, $N_l = 5$, $\theta = [87.7076^\circ, 93.4398^\circ, 154.1581^\circ]$, and $w = [0.02, -0.03, -0.45]$. (a)~$\|\bm{\psi}(\mathbf{Q}, w)\|_F$ versus $w$; (b)~$R(w)$ versus $w$.}
\label{dual_poly_eg}
\end{figure}

Now, suppose there exists a matrix $\mathbf{P}_0 \in \mathbb{C}^{N \times N}$ such that the constraints in \eqref{ssdp} hold. We must argue that $R(w) \geq 0$ and therefore $\|\mathcal{Q}\|^{*}_{\mathcal{A}} \leq 1$ for all $w$. Consider the expression $\mathbf{z}^H \mathbf{P}_0 \mathbf{z}$ and note that
\begin{equation}
\mathbf{z}^H \mathbf{P}_0 \mathbf{z} = \mathrm{Tr}(\mathbf{z} \mathbf{z}^H \mathbf{P}_0) = \sum_{k = -(N-1)}^{N-1} r_k z^{-k}
\end{equation}
where $r_k = \sum_{i = 1}^{N - k} \mathbf{P}_0(i, i + k)$ for $k \geq 0$ and $r_k = r^*_{-k}$ for $k < 0$. Since $\sum_{i = 1}^{N-k} \mathbf{P}_0(i, i+k) = \delta_k$ holds, we can conclude that $\mathbf{z}^H \mathbf{P}_0 \mathbf{z} = z^0 = 1$. Define $\mathbf{P}_1 := \sum_{f = 1}^{N_F} \widetilde{\mathbf{Q}}_f \widetilde{\mathbf{Q}}_f^H = \widetilde{\mathbf{Q}} \widetilde{\mathbf{Q}}^H$ and substitute this fact into $R(w)$. We have 
\begin{equation}
    R(w) = \mathbf{z}^H \mathbf{P}_0 \mathbf{z} - \mathbf{z}^H \mathbf{P}_1 \mathbf{z} = \mathbf{z}^H (\mathbf{P}_0 - \mathbf{P}_1) \mathbf{z}.
\end{equation}
Since $\left[                 
  \begin{array}{cc}   
    \mathbf{P}_0 & \widetilde{\mathbf{Q}} \\  
    \widetilde{\mathbf{Q}}^H & \mathbf{I}_{N_lN_F} \\  
  \end{array}
\right]  \succeq 0$, its Schur complement $\mathbf{P}_0 - \widetilde{\mathbf{Q}}\mathbf{I}_{N_F}^{-1} \widetilde{\mathbf{Q}}^H = \mathbf{P}_0 - \mathbf{P}_1 \succeq  0$, and so $R(w) \geq 0 $ for all $w \in [-1/2, 1/2]$.

Next, suppose $R(w) \geq 0 $ for all $w \in [-1/2, 1/2]$. We need to argue that there exists a matrix $\mathbf{P}_0 \in \mathbb{C}^{N \times N} \succeq 0$ such that the constraints in \eqref{ssdp} hold. Since $R(w) \leq 0$, $1 \geq \mathbf{z}^H \mathbf{P}_1 \mathbf{z}$, where $\mathbf{P}_1 := \sum_{f = 1}^{N_F} \widetilde{\mathbf{Q}}_f \widetilde{\mathbf{Q}}_f^H = \widetilde{\mathbf{Q}} \widetilde{\mathbf{Q}}^H$. From \cite[Lemma 4.25]{dumitrescu2017positive} and the fact that $1$ and $\mathbf{z}^H \mathbf{P}_1 \mathbf{z}$ are both univariate trigonometric polynomials, it follows that there exists $\mathbf{P}_0 \succeq \mathbf{P}_1 $ such that $1 = \mathbf{z}^H\mathbf{P}_0 \mathbf{z}$ and $\sum_{i = 1}^{N-k} \mathbf{P}_0(i, i+k) = \delta_k$ hold. The matrix $\left[                 
  \begin{array}{cc}   
    \mathbf{P}_0 & \widetilde{\mathbf{Q}} \\  
    \widetilde{\mathbf{Q}}^H & \mathbf{I}_{N_lN_F} \\  
  \end{array}
\right] $ has Schur complement $\mathbf{P}_0 - \widetilde{\mathbf{Q}}\mathbf{I}_{N_F}^{-1} \widetilde{\mathbf{Q}}^H  = \mathbf{P}_0 - \mathbf{P}_1 \succeq 0$, and therefore this matrix is positive semi-definite. This concludes the proof. $\hfill\square$

\bibliographystyle{IEEEtran}
\bibliography{strings}

\begin{thebibliography}{10}
\providecommand{\url}[1]{#1}
\csname url@samestyle\endcsname
\providecommand{\newblock}{\relax}
\providecommand{\bibinfo}[2]{#2}
\providecommand{\BIBentrySTDinterwordspacing}{\spaceskip=0pt\relax}
\providecommand{\BIBentryALTinterwordstretchfactor}{4}
\providecommand{\BIBentryALTinterwordspacing}{\spaceskip=\fontdimen2\font plus
\BIBentryALTinterwordstretchfactor\fontdimen3\font minus \fontdimen4\font\relax}
\providecommand{\BIBforeignlanguage}[2]{{%
\expandafter\ifx\csname l@#1\endcsname\relax
\typeout{** WARNING: IEEEtran.bst: No hyphenation pattern has been}%
\typeout{** loaded for the language `#1'. Using the pattern for}%
\typeout{** the default language instead.}%
\else
\language=\csname l@#1\endcsname
\fi
#2}}
\providecommand{\BIBdecl}{\relax}
\BIBdecl

\bibitem{van2002optimum}
H.~L. Van~Trees, \emph{Optimum array processing: Part IV of detection, estimation, and modulation theory}.\hskip 1em plus 0.5em minus 0.4em\relax John Wiley \& Sons, 2002.

\bibitem{chen2021millidegree}
Y.~Chen, L.~Yan, C.~Han, and M.~Tao, ``Millidegree-level direction-of-arrival estimation and tracking for terahertz ultra-massive mimo systems,'' \emph{IEEE Trans. Wirel. Comm.}, vol.~21, no.~2, pp. 869--883, 2021.

\bibitem{vasanelli2020calibration}
C.~Vasanelli, F.~Roos, A.~Durr, J.~Schlichenmaier, P.~Hugler, B.~Meinecke, M.~Steiner, and C.~Waldschmidt, ``Calibration and direction-of-arrival estimation of millimeter-wave radars: A practical introduction,'' \emph{IEEE Antennas and Propag. Mag.}, vol.~62, no.~6, pp. 34--45, 2020.

\bibitem{schmidt1986multiple}
R.~Schmidt, ``Multiple emitter location and signal parameter estimation,'' \emph{IEEE Trans. Antennas Propag.}, vol.~34, no.~3, pp. 276--280, 1986.

\bibitem{roy1989esprit}
R.~Roy and T.~Kailath, ``Esprit-estimation of signal parameters via rotational invariance techniques,'' \emph{IEEE Trans. Acoust., Speech, and Signal Process.}, vol.~37, no.~7, pp. 984--995, 1989.

\bibitem{wax1984spatio}
M.~Wax, T.-J. Shan, and T.~Kailath, ``Spatio-temporal spectral analysis by eigenstructure methods,'' \emph{IEEE Trans. Acoust., Speech, Signal Process.}, vol.~32, no.~4, pp. 817--827, 1984.

\bibitem{antonello2019joint}
N.~Antonello, E.~De~Sena, M.~Moonen, P.~A. Naylor, and T.~van Waterschoot, ``Joint acoustic localization and dereverberation through plane wave decomposition and sparse regularization,'' \emph{IEEE/ACM Trans. Audio, Speech, Lang. Process.}, vol.~27, no.~12, pp. 1893--1905, 2019.

\bibitem{nannuru2019sparse}
S.~Nannuru, K.~L. Gemba, P.~Gerstoft, W.~S. Hodgkiss, and C.~F. Mecklenbr{\"a}uker, ``Sparse {B}ayesian learning with multiple dictionaries,'' \emph{Signal Process.}, vol. 159, pp. 159--170, 2019.

\bibitem{gemba2019}
K.~L. Gemba, S.~Nannuru, and P.~Gerstoft, ``Robust ocean acoustic localization with sparse {B}ayesian learning,'' \emph{IEEE J. Sel. Topics Signal Process.}, vol.~13, no.~1, pp. 49--60, 2019.

\bibitem{wu2023gridless}
Y.~Wu, M.~B. Wakin, and P.~Gerstoft, ``Gridless {DOA} estimation with multiple frequencies,'' \emph{IEEE Trans. Signal Process.}, vol.~71, pp. 417--432, 2023.

\bibitem{wang1985coherent}
H.~Wang and M.~Kaveh, ``Coherent signal-subspace processing for the detection and estimation of angles of arrival of multiple wideband sources,'' \emph{IEEE Trans. Acoust., Speech, Signal Process.}, vol.~33, no.~4, pp. 823--831, 1985.

\bibitem{buckley1988broad}
K.~M. Buckley and L.~J. Griffiths, ``Broad-band signal-subspace spatial-spectrum ({BASS-ALE}) estimation,'' \emph{IEEE Trans. Acoust., Speech, Signal Process.}, vol.~36, no.~7, pp. 953--964, 1988.

\bibitem{di2001waves}
E.~D. Di~Claudio and R.~Parisi, ``{WAVES}: Weighted average of signal subspaces for robust wideband direction finding,'' \emph{IEEE Trans. Signal Process.}, vol.~49, no.~10, pp. 2179--2191, 2001.

\bibitem{yoon2006tops}
Y.-S. Yoon, L.~M. Kaplan, and J.~H. McClellan, ``{TOPS}: New {DOA} estimator for wideband signals,'' \emph{IEEE Trans. Signal Process.}, vol.~54, no.~6, pp. 1977--1989, 2006.

\bibitem{tang2011aliasing}
Z.~Tang, G.~Blacquiere, and G.~Leus, ``Aliasing-free wideband beamforming using sparse signal representation,'' \emph{IEEE Trans. Signal Process.}, vol.~59, no.~7, pp. 3464--3469, 2011.

\bibitem{gemba2017}
K.~L. Gemba, S.~Nannuru, P.~Gerstoft, and W.~S. Hodgkiss, ``Multi-frequency sparse {B}ayesian learning for robust matched field processing,'' \emph{J. Acoust. Soc. Am.}, vol. 141, no.~5, pp. 3411--3420, 2017.

\bibitem{zhang2013wideband}
J.~Zhang, N.~Hu, M.~Bao, X.~Li, and W.~He, ``Wideband {DOA} estimation based on block {FOCUSS} with limited samples,'' in \emph{IEEE GlobalSIP}, 2013, pp. 634--637.

\bibitem{wang2015novel}
L.~Wang, L.~Zhao, G.~Bi, C.~Wan, L.~Zhang, and H.~Zhang, ``Novel wideband {DOA} estimation based on sparse {B}ayesian learning with {D}irichlet process priors,'' \emph{IEEE Trans. Signal Process.}, vol.~64, no.~2, pp. 275--289, 2015.

\bibitem{liu2011broadband}
C.~Liu, Y.~V. Zakharov, and T.~Chen, ``Broadband underwater localization of multiple sources using basis pursuit denoising,'' \emph{IEEE Trans. Signal Process.}, vol.~60, no.~4, pp. 1708--1717, 2011.

\bibitem{zhang2021enhanced}
S.~Zhang, A.~Ahmed, Y.~D. Zhang, and S.~Sun, ``Enhanced {DOA} estimation exploiting multi-frequency sparse array,'' \emph{IEEE Trans. Signal Process.}, vol.~69, pp. 5935--5946, 2021.

\bibitem{chandrasekaran2012convex}
V.~Chandrasekaran, B.~Recht, P.~A. Parrilo, and A.~S. Willsky, ``The convex geometry of linear inverse problems,'' \emph{Found. Comput. Math.}, vol.~12, no.~6, pp. 805--849, 2012.

\bibitem{candes2014towards}
E.~J. Cand{\`e}s and C.~Fernandez-Granda, ``Towards a mathematical theory of super-resolution,'' \emph{Commun. Pure Appl. Math.}, vol.~67, no.~6, pp. 906--956, 2014.

\bibitem{tang2013compressed}
G.~Tang, B.~N. Bhaskar, P.~Shah, and B.~Recht, ``Compressed sensing off the grid,'' \emph{IEEE Trans. Inf. Theory}, vol.~59, no.~11, pp. 7465--7490, 2013.

\bibitem{li2015off}
Y.~Li and Y.~Chi, ``Off-the-grid line spectrum denoising and estimation with multiple measurement vectors,'' \emph{IEEE Trans. Signal Process.}, vol.~64, no.~5, pp. 1257--1269, 2015.

\bibitem{yang2016exact}
Z.~Yang and L.~Xie, ``Exact joint sparse frequency recovery via optimization methods,'' \emph{IEEE Trans. Signal Process.}, vol.~64, no.~19, pp. 5145--5157, 2016.

\bibitem{yang2018sample}
Z.~Yang, J.~Tang, Y.~C. Eldar, and L.~Xie, ``On the sample complexity of multichannel frequency estimation via convex optimization,'' \emph{IEEE Trans. Inf. Theory}, vol.~65, no.~4, pp. 2302--2315, 2018.

\bibitem{wagner2021}
M.~Wagner, Y.~Park, and P.~Gerstoft, ``Gridless {DOA} estimation and root-{MUSIC} for non-uniform linear arrays,'' \emph{IEEE Trans.\ Signal Process.}, vol.~69, pp. 2144--2157, 2021.

\bibitem{wu2022gridlessicassp}
Y.~Wu, M.~B. Wakin, and P.~Gerstoft, ``Gridless {DOA} estimation under the multi-frequency model,'' in \emph{IEEE ICASSP}, 2022, pp. 5982--5986.

\bibitem{jiang2020gridless}
Y.~Jiang, D.~Li, X.~Wu, and W.-P. Zhu, ``A gridless wideband {DOA} estimation based on atomic norm minimization,'' in \emph{IEEE Sensor Array and Multichannel Signal Processing Workshop (SAM)}.\hskip 1em plus 0.5em minus 0.4em\relax IEEE, 2020, pp. 1--5.

\bibitem{li2018atomic}
S.~Li, D.~Yang, G.~Tang, and M.~B. Wakin, ``Atomic norm minimization for modal analysis from random and compressed samples,'' \emph{IEEE Trans. Signal Process.}, vol.~66, no.~7, pp. 1817--1831, 2018.

\bibitem{chi2020harnessing}
Y.~Chi and M.~F. Da~Costa, ``Harnessing sparsity over the continuum: Atomic norm minimization for superresolution,'' \emph{IEEE Signal Process. Mag.}, vol.~37, no.~2, pp. 39--57, 2020.

\bibitem{chi2016guaranteed}
Y.~Chi, ``Guaranteed blind sparse spikes deconvolution via lifting and convex optimization,'' \emph{IEEE J. Sel. Topics Signal Process.}, vol.~10, no.~4, pp. 782--794, 2016.

\bibitem{ling2015self}
S.~Ling and T.~Strohmer, ``Self-calibration and biconvex compressive sensing,'' \emph{Inverse Problems}, vol.~31, no.~11, p. 115002, 2015.

\bibitem{moffet1968minimum}
A.~Moffet, ``Minimum-redundancy linear arrays,'' \emph{IEEE Trans. Antennas Propag.}, vol.~16, no.~2, pp. 172--175, 1968.

\bibitem{bloom1977applications}
G.~S. Bloom and S.~W. Golomb, ``Applications of numbered undirected graphs,'' \emph{Proceedings of the IEEE}, vol.~65, no.~4, pp. 562--570, 1977.

\bibitem{vaidyanathan2010sparse}
P.~P. Vaidyanathan and P.~Pal, ``Sparse sensing with co-prime samplers and arrays,'' \emph{IEEE Trans.\ Signal Process.}, vol.~59, no.~2, pp. 573--586, 2010.

\bibitem{pal2010nested}
P.~Pal and P.~P. Vaidyanathan, ``Nested arrays: A novel approach to array processing with enhanced degrees of freedom,'' \emph{IEEE Trans. Signal Process.}, vol.~58, no.~8, pp. 4167--4181, 2010.

\bibitem{wang2016coarrays}
M.~Wang and A.~Nehorai, ``Coarrays, music, and the {C}ram{\'e}r--{R}ao bound,'' \emph{IEEE Trans. Signal Process.}, vol.~65, no.~4, pp. 933--946, 2016.

\bibitem{dogan1995applications}
M.~C. Dogan and J.~M. Mendel, ``Applications of cumulants to array processing. i. aperture extension and array calibration,'' \emph{IEEE Trans. Signal Process.}, vol.~43, no.~5, pp. 1200--1216, 1995.

\bibitem{chevalier2005virtual}
P.~Chevalier, L.~Albera, A.~Ferr{\'e}ol, and P.~Comon, ``On the virtual array concept for higher order array processing,'' \emph{IEEE Trans. Signal Process.}, vol.~53, no.~4, pp. 1254--1271, 2005.

\bibitem{ma2009doa}
W.-K. Ma, T.-H. Hsieh, and C.-Y. Chi, ``{DOA} estimation of quasi-stationary signals via {K}hatri-{R}ao subspace,'' in \emph{IEEE ICASSP}, 2009, pp. 2165--2168.

\bibitem{grant2014cvx}
M.~Grant and S.~Boyd, ``C{VX}: Matlab software for disciplined convex programming, version 2.1,'' 2014.

\bibitem{xenaki2015}
A.~Xenaki and P.~Gerstoft, ``Grid-free compressive beamforming,'' \emph{J. Acoust. Soc. Am.}, vol. 137, pp. 1923--1935, 2015.

\bibitem{gerstoft2016}
P.~Gerstoft, C.~F. Mecklenbr{\"a}uker, A.~Xenaki, and S.~Nannuru, ``Multisnapshot sparse {B}ayesian learning for {DOA},'' \emph{IEEE Signal Process. Lett.}, vol.~23, no.~10, pp. 1469--1473, 2016.

\bibitem{rao1989performance}
B.~D. Rao and K.~S. Hari, ``Performance analysis of root-music,'' \emph{IEEE Trans. Acoust., Speech, Signal Process.}, vol.~37, no.~12, pp. 1939--1949, 1989.

\bibitem{liang2020review}
Y.~Liang, W.~Liu, Q.~Shen, W.~Cui, and S.~Wu, ``A review of closed-form {C}ram{\'e}r-{R}ao bounds for {DOA} estimation in the presence of gaussian noise under a unified framework,'' \emph{IEEE Access}, vol.~8, pp. 175\,101--175\,124, 2020.

\bibitem{liang2021cramer}
Y.~Liang, W.~Cui, Q.~Shen, W.~Liu, and H.~Wu, ``Cram{\'e}r-rao bound for {DOA} estimation exploiting multiple frequency pairs,'' \emph{IEEE Signal Process. Lett.}, vol.~28, pp. 1210--1214, 2021.

\bibitem{yang2018sparse}
Z.~Yang, J.~Li, P.~Stoica, and L.~Xie, ``Sparse methods for direction-of-arrival estimation,'' in \emph{Academic Press Library in Signal Processing, Volume 7}.\hskip 1em plus 0.5em minus 0.4em\relax Elsevier, 2018, pp. 509--581.

\bibitem{zhou2018direction}
C.~Zhou, Y.~Gu, X.~Fan, Z.~Shi, G.~Mao, and Y.~D. Zhang, ``Direction-of-arrival estimation for coprime array via virtual array interpolation,'' \emph{IEEE Trans.\ Signal Process.}, vol.~66, no.~22, pp. 5956--5971, 2018.

\bibitem{qin2017doa}
S.~Qin, Y.~D. Zhang, M.~G. Amin, and B.~Himed, ``{DOA} estimation exploiting a uniform linear array with multiple co-prime frequencies,'' \emph{Signal Process.}, vol. 130, pp. 37--46, 2017.

\bibitem{liu2021rank}
S.~Liu, Z.~Mao, Y.~D. Zhang, and Y.~Huang, ``Rank minimization-based {T}oeplitz reconstruction for {DoA} estimation using coprime array,'' \emph{IEEE Comm. Lett.}, vol.~25, no.~7, pp. 2265--2269, 2021.

\bibitem{dumitrescu2017positive}
B.~Dumitrescu, \emph{Positive Trigonometric Polynomials and Signal Processing Applications}.\hskip 1em plus 0.5em minus 0.4em\relax Springer, 2017, vol. 103.

\end{thebibliography}

\ifCLASSOPTIONcaptionsoff
  \newpage
\fi
\end{document}